\documentclass[aps,prb,notitlepage,twocolumn,superscriptaddress]{revtex4-2}

\usepackage{amsmath}
\usepackage{amssymb}
\usepackage{graphicx}
\usepackage{epstopdf}
\usepackage[colorlinks=true]{hyperref}
\usepackage{float}

\newcommand{\sech}{\mathrm{sech}}
\newcommand{\sgn}{\mathrm{sgn}}
\newcommand{\Tr}{\mathrm{Tr}}

\begin{document}
\title{Current-driven motion of magnetic topological defects in ferromagnetic superconductors}

\date{\today}

\author{Se Kwon Kim}
\email{sekwonkim@kaist.ac.kr}
\affiliation{Department of Physics, Korea Advanced Institute of Science and Technology, Daejeon 34141, Republic of Korea}
\author{Suk Bum Chung}
\email{sbchung0@uos.ac.kr}
\affiliation{Department of Physics and Natural Science Research Institute, University of Seoul, Seoul 02504, Republic of Korea}
\affiliation{School of Physics, Korea Institute for Advanced Study, Seoul 02455, Korea}

\begin{abstract}
Recent years have seen a number of instances where magnetism and superconductivity intrinsically coexist. Our focus is on the case where spin-triplet superconductivity arises out of ferromagnetism, and we make a hydrodynamic analysis of the effect of a charge supercurrent on magnetic topological defects like domain walls and merons. We find that the emergent electromagnetic field that arises out of the superconducting order parameter provides a description for not only the physical quantities such as the local energy flux density and the interaction between current and defects but also the energy dissipation through magnetic dynamics of the Gilbert damping, which becomes more prominent compared to the normal state as superconductivity attenuates the energy dissipation through the charge sector. In particular, we reveal that the current-induced dynamics of domain walls and merons in the presence of the Gilbert damping give rise to the nonsingular $4\pi$ and $2\pi$ phase slips, respectively, revealing the intertwined dynamics of spin and charge degrees of freedom in ferromagnetic superconductors. 
\end{abstract}

\maketitle

\section{Introduction}

While magnetism has traditionally been regarded as inimical to superconductivity, recent years have seen observation of ferromagnetism and superconductivity coexisting or cooperating in varieties of materials which includes uranium heavy-fermion compounds~\cite{SaxenaNature2000, AokiNature2001, AokiJPSJ2019} and two-dimensional moir\'e materials such as twisted bilayer graphene~\cite{ChenNP2019, SharpeScience2019, ZondinerNature2020}. It has been known that such coexistence can be naturally accommodated by the Cooper pairing of spin-polarized electrons~\cite{Vollhardt2013helium}. In such cases, it is natural to question 
what effect, if any, ferromagnetism may have on superconductivity and vice versa. 

It is well established in magnetism and spintronics that the current-induced motions of spin textures such as domain walls in magnetic metals give rise to the spin and energy dissipation into the baths of quasiparticles or phonons, commonly known as the Gilbert damping~\cite{GilbertIEEE2004, TserkovnyakJMMM2008}. The conservation of energy dictates that the dissipated energy should be externally supplied by the input power. 
In the case of normal metals, however, resistivity-induced energy dissipation is present regardless of the presence or the absence of any spin textures. Hence the Gilbert damping gives rise to only an additional term in the energy dissipation and, in this sense, its presence can be difficult to confirm solely through charge transport.

Charge transport detection of the Gilbert damping in ferromagnetic superconductors may be more straightforward despite involving 
a feature unconventional for superconductors. 
To maintain a steady-state motion of spin textures in the presence of the Gilbert damping, a ferromagnetic superconductor needs the finite input power that goes out of the superconductor {\it solely} in the form of the Gilbert damping. This indicates 
voltage 
arising inside the superconductor in the direction of the current by the dynamics of spin textures. The mechanisms by which a superconductor acquires a finite voltage difference between two points is referred to as phase slips~\cite{HalperinJLTP1979, Tinkham2004}. In conventional superconductors, these phase slips generally accompany the singularities, {\it i.e.} the vanishing of the order parameter at a certain time during the phase slips. 

\begin{figure}
\includegraphics[width=\columnwidth]{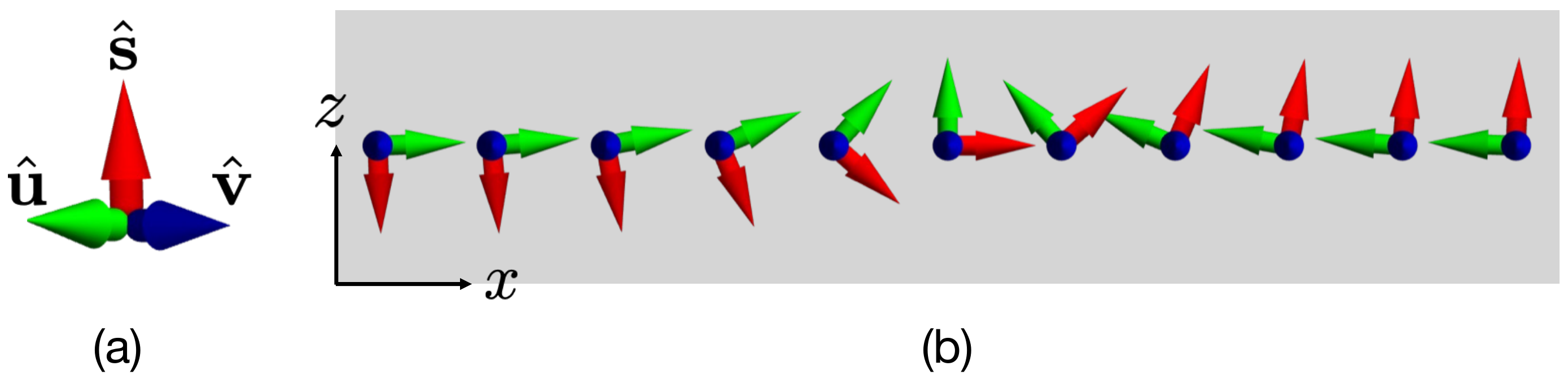}
\caption{(a) The illustration of the mutually orthogonal unit vectors $\hat{\mathbf{s}}, \hat{\mathbf{u}}$, and $\hat{\mathbf{v}}$ that describe the directional degrees of freedom of the order parameter of a ferromagnetic superconductor. (b) The configuration of the triad \{$\hat{\mathbf{s}}, \hat{\mathbf{u}}, \hat{\mathbf{v}}$\} for a domain wall in a ferromagnetic superconductor with easy-axis spin anisotropy along the $z$ direction.}
\label{fig:fig1}
\end{figure}

In this paper, we show that this is not necessarily the case for ferromagnetic superconductors by using the concrete example of the current-induced motions of two types of magnetic defects, domain walls and merons, which are schematically illustrated in Fig.~\ref{fig:fig1}(b) and Fig.~\ref{fig:v}, respectively. To this end, we begin by examining the order parameter of the spin-polarized superconductor and show how the Cooper pair spin rotation around the spin polarization direction is actually equivalent to the twisting of the overall phase. This gives rise to a channel for the interaction between ferromagnetism and superconductivity, namely the coupling of Cooper pairs to the effective gauge field arising from spin texture~\cite{Vollhardt2013helium, CornfeldPRR2021, PoniatowskiPRL2022}. 

We then proceed to show how such formalism can be used to obtain the current-induced motion of topological spin defects such as a domain wall and a meron in presence of a background superflow. First, for a domain wall, we show that the current-induced motion of domain walls in the presence of the Gilbert damping accompanies the precessional dynamics of the local spin polarization and this in turn gives rise to the nonsingular $4 \pi$ phase slips through the generation of an emergent electric field. The induced phase slip opens a channel through which the ferromagnetic superconductor can acquire input power, which is shown to be dissipated by the spin dynamics entirely via the Gilbert damping. Also, a current-induced motion of a meron is shown to give rise to the nonsingular $2 \pi$ phase slips perpendicular to its motion, engendering a channel for the input power that is dissipated via the Gilbert damping. The generation of the $2 \pi$ phase slips can be understood from the emergent electromagnetic field associated with the meron dynamics. For ferromagnetic metals, the emergent electromagnetic fields associated with spin textures and their dynamics have been discussed theoretically~\cite{VolovikJPC1987, WongPRB2009, ZangPRL2011, NagaosaNN2013} and confirmed experimentally~\cite{YangPRL2009, NeubauerPRL2009, YangPRB2010, BisigPRL2016}. However, their manifestations in the dynamics of magnetic defects in ferromagnetic superconductors and the resultant nonsingular phase slips have not been discussed yet. Our work reveals that the current-induced dynamics of magnetic defects exemplify the intertwined dynamics of spin and charge degrees of freedom in ferromagnetic superconductors, where the emergent electromagnetic fields play crucial roles.

The paper is organized as follows. The general formalism for the order parameter and its dynamics of ferromagnetic superconductors is developed phenomenologically in Sec.~\ref{sec:formalism}. The current-induced dynamics of a domain wall and its relation to the nonsingular $4 \pi$ phase slips are discussed in Sec.~\ref{sec:dw}. Section~\ref{sec:meron} concerns the current-induced dynamics of a meron and its relation to the nonsingular $2 \pi$ phase slips. We conclude the paper in Sec.~\ref{sec:discussion} with discussions.

\section{General formalism}
\label{sec:formalism}

\subsection{Order parameter}

The order parameter of a fully spin-polarized triplet superconductor provides a starting point for understanding how superconductivity and magnetism are intertwined through the emergent gauge field. In the d-vector formalism defined by $i(\mathbf{d}\cdot\boldsymbol{\sigma}\sigma^y)_{s,s'} \equiv \Delta_{s,s'}$, it is given by~\cite{Vollhardt2013helium, CornfeldPRR2021, PoniatowskiPRL2022}
\begin{equation}
\mathbf{d} = \frac{\sqrt{\rho}}{2} e^{i \phi} (\hat{\mathbf{u}} + i \hat{\mathbf{v}}) = \sqrt{\frac{\rho}{2}} \frac{e^{i \phi} (\hat{\mathbf{u}} + i \hat{\mathbf{v}})}{\sqrt{2}} \equiv \sqrt{\frac{\rho}{2}} \hat{\mathbf{d}} \, ,
\label{EQ:OP}
\end{equation}
where $\rho = 2 \mathbf{d}^* \cdot \mathbf{d}$ is the number density of the Cooper pairs, $\hat{\mathbf{u}}$ and $\hat{\mathbf{v}}$ are perpendicular unit vectors, and $\hat{\mathbf{d}}^* \cdot \hat{\mathbf{d}} = 1$; the simplest example would be $\hat{\mathbf{u}}=\hat{\mathbf{x}}, \hat{\mathbf{v}}=\hat{\mathbf{y}}$ which gives $\Delta_{s,s'}=0$ except for $\Delta_{\uparrow\uparrow}$ (see Appendix~\ref{app:op} for the details). There is an ambiguity here in defining $\phi$ as the above order parameter remains invariant under the following simultaneous change of $\phi$ and $\hat{\mathbf{u}}$ and $\hat{\mathbf{v}}$:
\begin{equation}
e^{i \phi} (\hat{\mathbf{u}} + i \hat{\mathbf{v}}) = e^{i (\phi + \delta \phi)} \left[ e^{- i \delta \phi} (\hat{\mathbf{u}} + i \hat{\mathbf{v}}) \right] \equiv e^{i (\phi + \delta \phi)} (\hat{\mathbf{u}}' + i \hat{\mathbf{v}}') \, ,
\label{EQ:OPsymm}
\end{equation}
where $\hat{\mathbf{u}}', \hat{\mathbf{v}}'$ are obtained by rotating $\hat{\mathbf{u}}, \hat{\mathbf{v}}$ by $+\delta \phi$ around $\hat{\mathbf{u}} \times \hat{\mathbf{v}}$. As the spin density in units of $\hbar$ can be written as
\begin{equation}
\mathbf{s} = 2 i \mathbf{d} \times \mathbf{d}^* = \rho \hat{\mathbf{u}} \times \hat{\mathbf{v}} \equiv \rho \hat{\mathbf{s}} \, ,
\label{EQ:CooperSpin}
\end{equation}
Eq.~\eqref{EQ:OPsymm} denotes the ${\rm U}(1)_{\phi + s}$ order parameter redundancy   \cite{Vollhardt2013helium, CornfeldPRR2021}, {\it i.e.} the invariance of the order parameter when the angle of the spin rotation around $\hat{\mathbf{s}}$ equals the change in $\phi$. Such redundancy implies the existence of an effective gauge field arising from the spin degrees of freedom. See Fig.~\ref{fig:fig1}(a) for the illustration of the three mutually orthogonal unit vectors $\hat{\mathbf{s}}, \hat{\mathbf{u}}$, and $\hat{\mathbf{v}}$, which are depicted by red, green, and blue arrows, respectively.

For deriving the vector potential and magnetostatics of this effective gauge field, the above order parameter suffices. From the spin rotation angle around $\hat{\bf s}$ defined as $\alpha$, the effective vector gauge can be written as~\cite{VolovikJPC1987}
\begin{equation}
a_i \equiv \frac{\hbar}{q} \partial_i \alpha = \frac{\hbar}{q} \hat{\mathbf{s}} \cdot (\hat{\mathbf{u}} \times \partial_i \hat{\mathbf{u}}) \, ;
\label{EQ:eGauge}
\end{equation}
hence the emergent gauge is a direct consequence of the ${\rm U}(1)_{\phi +s}$ order parameter redundancy of Eq.~\eqref{EQ:OPsymm}. Indeed, the emergent gauge field of spatial curvature in the chiral superconductor has been attributed to the analogous order parameter redundancy there \cite{Cross1975, Mermin1978, QDJiang2020, QDJiang2022}. Here, while we have kept the charge, $q$, generic, $q = - 2 e < 0$ holds in superconductors. From this emergent vector potential, it is straightforward to obtain the emergent magnetic field,
\begin{equation}
b_i = \epsilon_{ijk} \partial_j a_k = - \frac{\hbar \epsilon_{ijk}}{2q} \hat{\mathbf{s}} \cdot (\partial_j \hat{\mathbf{s}} \times \partial_k \hat{\mathbf{s}}) \, ;
\end{equation}
note that this is in the same form as the well-known Mermin-Ho relation between the orbital angular momentum texture and superfluid velocity in the 3He-A superfluid~\cite{MerminHo1976, Volovik2003}. Yet this discussion does not include any dynamics, for which we shall adopt a two-step approach of first formulating the simplest free energy for the order parameter [Eq.~\eqref{EQ:OPsymm}] and then use its Lagrangian to obtain the equations of motion.

\subsection{Free energy}

Given that we seek results relevant to wide-ranging superconductors whose common attributes may not extend beyond the spin-polarized Cooper pairing \cite{SaxenaNature2000, AokiNature2001, AokiJPSJ2019, ChenNP2019, SharpeScience2019, ZondinerNature2020} we will consider for our free energy the simplest minimal model that includes the spin anisotropy and the Zeeman coupling:
\begin{equation}
F[\mathbf{d}] = \int dV \mathcal{F}'_0[\mathbf{d}] + \int dV \frac{U}{2} \left(2 |\mathbf{d}|^2 - \rho_0 \right)^2\, ,
\end{equation}
where
$$
\mathcal{F}'_0 \!=\! \frac{A'_d}{2} |(\boldsymbol{\nabla} - i \frac{q}{\hbar} \mathbf{A}) \mathbf{d}|^2 + \rho\!\left[\frac{A'_s}{2} |\boldsymbol{\nabla} \hat{\mathbf{s}}|^2 \!-\! \frac{D}{2} (\hat{\mathbf{s}} \!\cdot\! \hat{\mathbf{z}})^2 \!-\! H s_z \!+\! q V\right] \, ,
$$
where $A'_s$ represents the excess spin stiffness; note that, in contrast to previous analysis \cite{CornfeldPRR2021}, our treatment will encompass both the easy-axis anisotropy $D>0$ and the easy-plane anisotropy $D<0$. As we will focus on the cases where the fluctuation of the condensate density $\rho \equiv 2 |\mathbf{d}|^2$ is strongly suppressed, it is convenient to separately group together terms dependent on $\rho$ fluctuations~\cite{CornfeldPRR2021, PoniatowskiPRL2022}
$$
F = \int dV \rho \mathcal{F}_0 + \int dV  \left[\frac{A'_d}{16\rho}(\boldsymbol{\nabla}\rho)^2+\frac{U}{2}(\rho - \rho_0)^2\right] \, , 
$$
where 
$$
\mathcal{F}_0= \frac{A'_d}{2} |(\boldsymbol{\nabla} - i \frac{q}{\hbar} \mathbf{A}) \hat{\mathbf{d}}|^2 + \frac{A'_s}{2} |\boldsymbol{\nabla} \hat{\mathbf{s}}|^2 - \frac{D}{2} (\hat{\mathbf{s}} \cdot \hat{\mathbf{z}})^2 - H s_z + q V
$$ 
is the free energy density per unit density. The gauge transformation is implemented as
$$
\mathbf{A} \mapsto \mathbf{A} + \boldsymbol{\nabla} \Lambda \, , \quad \mathbf{d} \mapsto e^{i q \Lambda / \hbar} \mathbf{d} \, .
$$

The free energy can be recast into the following form:
\begin{widetext}
\begin{equation}
F = \int dV \rho \left\{ \frac{A_c}{2} \left[ \partial_i \phi - \frac{q}{\hbar} A_i - \hat{\mathbf{s}} \cdot (\hat{\mathbf{u}} \times \partial_i \hat{\mathbf{u}}) \right]^2 + \frac{A_s}{2 } (\partial_i \hat{\mathbf{s}})^2 - \frac{D}{2} (\hat{\mathbf{s}} \cdot \hat{\mathbf{z}})^2 - H s_z + q V \right\} + \int dV \left[\frac{A_c}{8\rho}(\boldsymbol{\nabla}\rho)^2+\frac{U}{2}(\rho - \rho_0)^2\right] \, ,
\end{equation}
\end{widetext}
where $A_c = A'_d$ and $A_s = A'_s + A'_d/2$. Here, $A_c$ and $A_s$ represent the charge stiffness and the spin stiffness, respectively. The similar expression without the second and the third terms can be found in Eq.~(2.3) and Eq.~(2.5) of Ref.~\cite{LiPRB2009}. 

Accordingly, the charge supercurrent density is modified to
\begin{equation}
J_i = - \frac{\delta F}{\delta A_i} = \frac{q}{\hbar} \rho A_c (\partial_i \phi - \frac{q}{\hbar} A_i - \frac{q}{\hbar} a_i) \, ,
\end{equation}
with the velocity field is given by
\begin{equation}
v_i = \frac{J_i}{q \rho} = \frac{A_c}{\hbar} (\partial_i \phi - \frac{q}{\hbar} A_i - \frac{q}{\hbar} a_i) \, .
\end{equation}
It satisfies the following equation (by assuming a nonsingular $\theta$):
\begin{equation}
\boldsymbol{\nabla} \times \mathbf{v} = - \frac{q A_c}{\hbar^2} (\mathbf{B} + \mathbf{b}) \, .
\end{equation}

The free energy formalism provides a convenient springboard for extending our analysis to dynamics as well. In particular, such analysis helps us understand how emergent electric field would arise, in analogy with the standard electrodynamics. This is accomplished by considering the Langrangian for this minimal model.

\subsection{Equations of motion}

For the dynamics analysis, we now obtain the classical equation of motion for both the charge and the spin component of the order parameter through considering the Lagrangian of the spin-polarized superconductor.
This can be written as
\begin{eqnarray}
L &=&\int dV 2 i \hbar \mathbf{d}^* \cdot \partial_t \mathbf{d} - F
\,  \nonumber\\
&=& - \int dV \hbar \rho [\partial_t \phi -\hat{\mathbf{s}} \cdot (\hat{\mathbf{u}} \times \partial_t \hat{\mathbf{u}})] - F \, .
\end{eqnarray}
The first term of the above Lagrangian arises from $2i\hbar {\bf d}^*$ being the conjugate variable to ${\bf d}$; the detailed derivation of its relation to $\rho, \phi, \hat{\mathbf{s}}$ can be found in Appendix~\ref{app:kin}. The low-energy dynamics of the order parameter can be described by the three Euler-Lagrange equations for $\phi, \rho$, and $\hat{\mathbf{s}}$. 

The equations for $\rho$ and $\phi$ are basically analogous to those of the conventional superconductors. The equation of motion for the density $\rho$,
\begin{eqnarray}
\dot{\rho} &=& - \frac{1}{\hbar} \partial_i \left\{ \rho A_c \left[ \partial_i \phi - \frac{q}{\hbar} A_i - \hat{\mathbf{s}} \cdot (\hat{\mathbf{u}} \times \partial_i \hat{\mathbf{u}}) \right] \right\} = - \frac{1}{q} \partial_i J_i \, , \nonumber\\
&=& - \boldsymbol{\nabla} \cdot (\rho \mathbf{v}) \,  
\end{eqnarray}
is obtained from $\delta L/\delta \phi = 0$ and is none other than the continuity equation for the Cooper pair density. Similarly, the equation of motion for the phase $\phi$
\begin{equation}
- \hbar [\partial_t \phi -\hat{\mathbf{s}} \cdot (\hat{\mathbf{u}} \times \partial_t \hat{\mathbf{u}})] = \mathcal{F}_0 + U (\rho - \rho_0)-\frac{A_c}{4\rho}\nabla^2 \rho\, 
\end{equation}
obtained from $\delta L/\delta \rho = 0$ (where only terms constant or linear in $\rho$ are retained) comes out to be the Josephson relation. We, however, want to obtain a hydrodynamic equation of motion for Cooper pairs, for which purpose we take the spatial derivative
of the Josephson relation:
\begin{widetext}
$$
- \hbar \left[\partial_t \partial_i \phi - \frac{q}{\hbar} (E_i + e_i) - \frac{q}{\hbar} \partial_t (A_i + a_i)\right] = \frac{\hbar^2}{A_c} \mathbf{v} \cdot \partial_i \mathbf{v} + \partial_i \left[ \rho'_e + U (\rho - \rho_0) -\frac{A_c}{4\rho}\nabla^2 \rho \right] \, ,
$$
where
\begin{equation}
e_i = - \frac{\hbar}{q} \hat{\mathbf{s}} \cdot (\partial_i \hat{\mathbf{s}} \times \partial_t \hat{\mathbf{s}}) \, 
\end{equation}
is the emergent electric field and
\begin{equation}
\rho'_e = A_s\frac{(\partial_i \hat{\mathbf{s}})^2}{2} - D\frac{s_z^2}{2} - H s_z
\end{equation}
is the magnetic energy density (per unit density). By using
$$
\mathbf{v} \cdot \partial_i \mathbf{v} = \frac{1}{2}  \partial_i (\mathbf{v}^2) = (\mathbf{v} \cdot \boldsymbol{\nabla}) v_i + \epsilon_{ijk} v_j (\boldsymbol{\nabla} \times \mathbf{v})_k = (\mathbf{v} \cdot \boldsymbol{\nabla}) v_i - \frac{q A_c}{\hbar^2} \epsilon_{ijk} v_j  (B_k + b_k) \, 
$$
and defining the material derivative $D_t \equiv \partial_t + \mathbf{v} \cdot \boldsymbol{\nabla}$ and the effective mass of a Cooper pair, $m \equiv \hbar^2 / A_c$, we obtain
\begin{equation}
m D_t \mathbf{v} = q (\mathbf{E} + \mathbf{e}) + q \mathbf{v} \times (\mathbf{B} + \mathbf{b}) - \partial_i \left[ \rho'_e + U (\rho - \rho_0) -\frac{A_c}{4\rho}\nabla^2 \rho \right] \, .
\end{equation}
\end{widetext}

The novelty in the ferromagnetic superconductor is the equation of motion for the spin direction $\hat{\mathbf{s}}$ that is 
derived from $\delta L/\delta \hat{\mathbf{s}}=0$ (see Appendix~\ref{app:eom} for details):
\begin{equation}
\hbar \rho \partial_t \hat{\mathbf{s}} = - \frac{\hbar J_i}{q} \partial_i \hat{\mathbf{s}} + \partial_i [\rho A_s (\hat{\mathbf{s}} \times \partial_i \hat{\mathbf{s}})] + \rho D (\hat{\mathbf{s}} \cdot \hat{\mathbf{z}}) \hat{\mathbf{s}} \times \hat{\mathbf{z}} + \rho H \hat{\mathbf{s}} \times \hat{\mathbf{z}} \, ,
\end{equation}
which is identical to the Landau-Lifshitz equation~\cite{LL5} augmented by the adiabatic spin-transfer torque~\cite{SlonczewskiJMMM1996, BergerPRB1996, TserkovnyakJMMM2008}. This can be also written as
$$
\hbar \rho D_t \hat{\mathbf{s}} = \partial_i [\rho A_s (\hat{\mathbf{s}} \times \partial_i \hat{\mathbf{s}})] + \rho D (\hat{\mathbf{s}} \cdot \hat{\mathbf{z}}) \hat{\mathbf{s}} \times \hat{\mathbf{z}} + \rho H \hat{\mathbf{s}} \times \hat{\mathbf{z}} \, .
$$
By using $\dot{\rho} = - \partial_i J_i / q$, we can obtain the spin continuity equation:
\begin{eqnarray}
\partial_t (\hbar \rho \hat{\mathbf{s}}) = - \partial_i \mathbf{J}^s_i + \rho D (\hat{\mathbf{s}} \cdot \hat{\mathbf{z}}) \hat{\mathbf{s}} \times \hat{\mathbf{z}} + \rho H \hat{\mathbf{s}} \times \hat{\mathbf{z}} \, ,
\end{eqnarray}
where
$$
\mathbf{J}^s_i = \frac{\hbar J_i}{q} \hat{\mathbf{s}} - \rho A_s (\hat{\mathbf{s}} \times \partial_i \hat{\mathbf{s}}) \, ,
$$
is the spin current density. The first term and the second term on the right-hand side are longitudinal spin currents proportional to the charge current and the transverse spin current that is carried by a spin texture, respectively.

A complete set of equations describing the hydrodynamics of a ferromagnetic superconductor in the absence of external fields ($\mathbf{E} = 0$ and $\mathbf{B} = 0$) can now be given; the analogous equations have been written down for a spinor BEC~\cite{BarnettPRB2009}. It is convenient to measure energy in the unit of the anisotropy energy absolute value $|D|$ and length in the unit derived from the combination of $|D|$ with the spin stiffness $A_s$, {\it i.e.}
$$
l = \sqrt{\frac{A_s}{|D|}} \, , \quad \epsilon = |D| \, .
$$
Also, we will use $\tilde{\rho} \equiv \rho / \rho_0, U \rho / D \equiv \eta$. This gives us the dimensionless equations
\begin{eqnarray}
- D_t \tilde{\rho} &=& \tilde{\rho} (\boldsymbol{\nabla} \cdot \mathbf{v}) \, , \nonumber\\
\tilde{m} (\boldsymbol{\nabla} \times \mathbf{v}) &=& - \mathbf{b} \, , \nonumber\\
\tilde{m} D_t \mathbf{v} &=& \mathbf{e} + \mathbf{v} \times \mathbf{b}\nonumber\\ 
&\,&- \partial_i [\rho_e \!+\! \eta (\tilde{\rho} - 1) -\nabla^2 \tilde{\rho}/4] \, , \nonumber\\
\tilde{\rho} D_t \hat{\mathbf{s}} &=& \partial_i [\tilde{\rho} (\hat{\mathbf{s}} \times \partial_i \hat{\mathbf{s}})] + \tilde{\rho}\nu (\hat{\mathbf{s}} \cdot \hat{\mathbf{z}}) \hat{\mathbf{s}} \times \hat{\mathbf{z}} + \tilde{\rho} h \hat{\mathbf{s}} \times \hat{\mathbf{z}} \, ,\nonumber\\
\label{EQ:hydro}
\end{eqnarray}
where $\nu \equiv {\rm sgn}(D)$,  $\tilde{m} \equiv A_s / A_c$ the dimensionless mass which is on the order of unity,  $h \equiv H / D$ the dimensionless external field, and 
$$
\rho_e = \frac{1}{2} (\partial_i \hat{\mathbf{s}})^2 - \nu \frac{1}{2} (\hat{\mathbf{s}} \cdot \hat{\mathbf{z}})^2 - hs_z
$$
the dimensionless magnetic energy density; for zero excess spin stiffness $\tilde{m} = 1/2$. The emergent electromagnetic fields are now re-defined as
$$
e_i = - \hat{\mathbf{s}} \cdot (\partial_i \hat{\mathbf{s}} \times \partial_t \hat{\mathbf{s}}) \, , \quad b_i = - \frac{\epsilon_{ijk}}{2} \hat{\mathbf{s}} \cdot (\partial_j \hat{\mathbf{s}} \times \partial_k \hat{\mathbf{s}}) \, ,
$$
where the charge $q$ is absorbed into the fields.

\subsection{Gilbert damping}

Due to the inevitable nonconservation of spin angular momentum in solids, it is reasonable to expect the damping of spin dynamics and the associated energy dissipation, which are not included in the hydrodynamics equations [Eq.~\eqref{EQ:hydro}], to play an important role in spin dynamics of the ferromagnetic superconductors as in any other solid-state systems. The spin sinks can be quasiparticles, phonons, and any other excitations that can possess angular momentum~\cite{TserkovnyakPRL2002, HickeyPRL2009, StarikovPRL2010, RuckriegelPRB2014}. The spin dissipation can be treated phenomenologically with the addition of the Gilbert damping term $\alpha \tilde{\rho} \hat{\mathbf{s}} \times \partial_t \hat{\mathbf{s}}$~\cite{GilbertIEEE2004} to the spin equation of motion in Eq.~\eqref{EQ:hydro},
\begin{equation}
\tilde{\rho} D_t \hat{\mathbf{s}} + \alpha \tilde{\rho} \hat{\mathbf{s}} \times \partial_t \hat{\mathbf{s}} = \partial_i [\tilde{\rho} (\hat{\mathbf{s}} \times \partial_i \hat{\mathbf{s}})] + \tilde{\rho} \nu (\hat{\mathbf{s}} \cdot \hat{\mathbf{z}}) \hat{\mathbf{s}} \times \hat{\mathbf{z}} + \tilde{\rho} h \hat{\mathbf{s}} \times \hat{\mathbf{z}} \, .
\label{EQ:GDamp}
\end{equation}
In the incompressible limit $\eta \rightarrow \infty$ where $\tilde{\rho}$ is uniform and constant, this gives us the energy continuity equation,
\begin{equation}
\label{eq:cont}
\partial_t \rho_e + \boldsymbol{\nabla} \cdot \mathbf{j}_e = - \mathbf{v} \cdot \mathbf{e} - \alpha (\partial_t \hat{\mathbf{s}})^2 \, ,
\end{equation}
where
\begin{equation}
\mathbf{j}_e = - \partial_t \hat{\mathbf{s}} \cdot \boldsymbol{\nabla} \hat{\mathbf{s}}
\end{equation}
is the magnetic energy flux density (per unit density). This continuity equation, which has been previously noted in literature~\cite{KimPRB2014, DasguptaPRB2018}, can be derived by taking the product of both sides of Eq.~\eqref{EQ:GDamp} with $\hat{\mathbf{s}} \times \partial_t \hat{\mathbf{s}}$. One can note that the first term on the right-hand side of Eq.~\eqref{eq:cont} ($- \mathbf{v} \cdot \mathbf{e}$) is the power dissipated (supplied) by the superflow $\mathbf{v}$ flowing parallel (antiparallel) to the direction of the emergent electric field $\mathbf{e}$, while the second term 
($- \alpha (\partial_t \hat{\mathbf{s}})^2$) is the energy dissipation through the Gilbert damping.

Equation~\eqref{eq:cont} implies that, in the incompressible limit, the energy dissipated by the Gilbert damping is equal to the work done by the emergent electric field when spin texture is transported without any distortion. This is because the total magnetic energy should be unchanged in this process and hence the left-hand side of Eq.~\eqref{eq:cont} integrated over the whole system should be zero.

\begin{figure*}
\includegraphics[width=1.9\columnwidth]{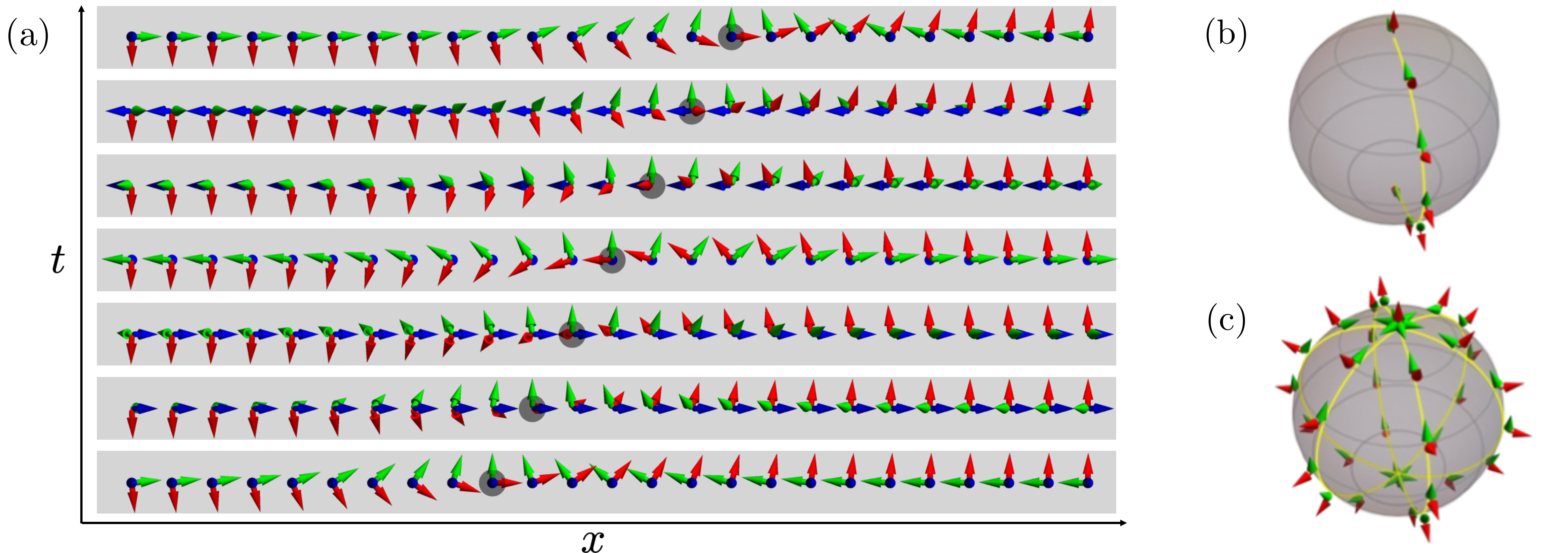}
\caption{(a) A series of snapshots of a precessing domain wall moving to the right, where $x$ and $t$ are the spatial and the temporal coordinates, respectively. The red arrows, blue arrows, and green arrows represent $\hat{\mathbf{s}}, \hat{\mathbf{u}}$, and $\hat{\mathbf{v}}$, respectively. The domain-wall position is denoted by the gray dot. The domain-wall angle, which is the azimuthal angle of $\hat{\mathbf{s}}$ at the center of the domain wall, changes from $\varphi_0 = 0$ to $\varphi_0 = - 2 \pi$ gradually with increasing time from bottom to top. On the left end, $\hat{\mathbf{u}}$ (green arrow) rotates by $- 2 \pi$ about $\hat{\mathbf{s}}$ (red arrow), whereas on the right end, $\hat{\mathbf{u}}$ rotates by $2 \pi$ about $\hat{\mathbf{s}}$. This process of a domain-wall precession can be considered as a nonsingular $4 \pi$ phase slip since these opposite $2 \pi$ rotations of $\hat{\mathbf{u}}$ around $\hat{\mathbf{s}}$ at the left and the right ends induces a finite voltage across the wire. (b) Mapping of the instantaneous configuration of the two vectors $\hat{\mathbf{s}}$ (red arrows) and $\hat{\mathbf{u}}$ (green arrows) onto the unit sphere with the normal vector identified with $\hat{\mathbf{s}}$. The yellow line represents a spatial dimension of the system. (c) Collection of the mapping of the configuration of $\hat{\mathbf{s}}$ and $\hat{\mathbf{u}}$ onto the unit sphere with $\hat{\mathbf{s}}$ identified with the normal vector for all the snapshots shown in (a). Note that the unit tangent vector field $\hat{\mathbf{u}}$ is not uniquely determined at the north and the south poles 
as dictated by the Poincar\'e-Hopf theorem~\cite{KamienRMP2002}. Rather, $\hat{\mathbf{u}}$ rotates once around $\hat{\mathbf{s}}$ counterclockwise (counterclockwise) at the north (south) pole as the domain wall completes one cycle of rotation, which is consistent with the Euler number 2 of the sphere. See the main text for further detailed discussions.}
\label{fig:fig2}
\end{figure*}

\section{Domain wall}
\label{sec:dw}

For the supercurrent-driven motion of topological defects, we first consider an easy-axis ferromagnetic superconductor with spin-anisotropy sign $\nu = \sgn(D) =+1$. A domain wall is a generically stable topological defect between two different ground states for easy-axis spin systems and intrinsically has no skyrmion density, {\it i.e.} no emergent magnetic field. Therefore for the rest of this section, we will take the Cooper pair velocity to be irrotational, {\it i.e.} $\boldsymbol{\nabla} \times {\bf v} = 0$ [Eq.~(\ref{EQ:hydro})]. In addition, we also set the applied magnetic field to be zero and hence $h=0$.

\subsection{Deriving dynamics from a static solution}

The solution for a domain-wall motion in the background of the constant and uniform background superflow can be straightforwardly constructed from the static domain wall solution in absence of any background superflow for the incompressible limit $\eta \to \infty$. For this case, the absence of the emergent magnetic field allows us to consider only the spin equation of motion~\cite{Takashima2017}
\begin{equation}
(\partial_t +\mathbf{v}\cdot \boldsymbol{\nabla} + \alpha \hat{\mathbf{s}} \times \partial_t) \hat{\mathbf{s}} = \partial_i [(\hat{\mathbf{s}} \times \partial_i \hat{\mathbf{s}})] + (\hat{\mathbf{s}} \cdot \hat{\mathbf{z}}) \hat{\mathbf{s}} \times \hat{\mathbf{z}} \, ;
\label{EQ:dw}
\end{equation}
from the hydrodynamic equations Eq.~\eqref{EQ:hydro}; others are either irrelevant due to ${\bf b}=0$ or merely provides the constraint $\tilde{\rho}=1$ in the incompressible limit. Given the intrinsically quasi-one-dimensional nature of the domain wall, we can set the boundary condition
$$
\hat{\mathbf{s}} (x \rightarrow \pm \infty) = \pm \hat{\mathbf{z}} \, ,
$$
for any domain-wall configuration. For the static domain wall at $x=0$ in absence of background superflow, the solution is given by the following Walker ansatz~\cite{SchryerJAP1974}:
\begin{equation}
\hat{\mathbf{s}}_0 = (\hat{\mathbf{x}} \cos \varphi_0 + \hat{\mathbf{y}} \sin \varphi_0) \sech x + Q  \hat{\mathbf{z}} \tanh x,
\label{EQ:dwStatic}
\end{equation} 
where $Q=\pm 1$ represents the domain wall type, satisfies the static domain-wall equation
\begin{equation}
0=\partial_x [(\hat{\mathbf{s}}_0 \times \partial_x \hat{\mathbf{s}}_0)] + (\hat{\mathbf{s}}_0 \cdot \hat{\mathbf{z}}) \hat{\mathbf{s}}_0 \times \hat{\mathbf{z}}\, ,
\label{EQ:dwStaticEq}
\end{equation}
derived from Eq.~\eqref{EQ:dw}, for an arbitrary domain-wall angle $\varphi_0$; it is important to note here that Eq.~\eqref{EQ:dwStaticEq} is sufficient as Eq.~\eqref{EQ:dwStatic} gives us ${\bf v} = 0$ and ${\bf b} = 0$ everywhere.

From Eq.~\eqref{EQ:dwStaticEq}, it can be shown that the general solution can be obtained by applying to the static solution both giving boost in the spatial direction and precession around the easy-axis:
\begin{equation}
\hat{\mathbf{s}} = (\hat{\mathbf{x}} \cos \Omega t + \hat{\mathbf{y}} \sin \Omega t) \sech (x-Vt) + Q  \hat{\mathbf{z}} \tanh (x-Vt)\, .
\label{EQ:dwGen}
\end{equation}
In deriving the domain-wall velocity $\hat{\mathbf{x}}V$ and the precession rate $\Omega$, it is convenient to note that Eq.~\eqref{EQ:dwGen} also satisfies Eq.~\eqref{EQ:dwStaticEq} as the latter equation involves no time derivatives. Also, as Eq.~\eqref{EQ:dwGen} is obtained from boost and precession, 
$$
\partial_t \hat{\mathbf{s}} = (- V\partial_x + \hat{\mathbf{z}}\Omega \times)  \hat{\mathbf{s}}\, .
$$
The velocity $V$ and the precession rate $\Omega$ therefore can be obtained from
\begin{equation}
[v\partial_x + (1+\alpha \hat{\mathbf{s}} \times)(-V\partial_x +  \hat{\mathbf{z}}\Omega \times)]\hat{\mathbf{s}}=0\, ,
\end{equation}
where we set the background superflow to be perpendicular to the domain wall without any loss of generality, $\mathbf{v} = v \hat{\mathbf{x}}$. By taking the scalar product of the above equation with $\hat{\mathbf{z}}$ and $\hat{\mathbf{z}}\times\hat{\mathbf{s}}$ we obtain
$$
v-V=Q\alpha\Omega\, , \quad \Omega = - Q \alpha V
$$
respectively, giving us
\begin{equation}
V = \frac{1}{1 + \alpha^2} v \, , \quad \Omega 
= - \frac{Q \alpha}{1 + \alpha^2} v \, ;
\end{equation}
Note that in absence of the Gilbert damping, $\alpha = 0$, there would have been no precession and the domain wall would have remained static with respect to the background superflow. See Fig.~\ref{fig:fig2} for the illustration of the domain-wall dynamics with precession.

From the above solution, it is straightforward to confirm that all work done by the emergent electric field is dissipated through the Gilbert damping. The work done by the emergent electric field is
\begin{equation}
- \mathbf{v} \cdot \mathbf{e} = v_j \hat{\mathbf{s}} \cdot (\partial_j \hat{\mathbf{s}} \times \partial_t \hat{\mathbf{s}}) = Q v \Omega [1-(\hat{\mathbf{s}} \cdot \hat{\mathbf{z}})^2] \, ,
\end{equation}
which gives the total energy input of
\begin{equation}
\label{eq:W}
W = \int dx (- \mathbf{v} \cdot \mathbf{e}) = 2 Q v \Omega \, .
\end{equation}
The energy dissipation rate per unit density is given by
\begin{equation}
\alpha (\partial_t \hat{\mathbf{s}})^2 = \alpha (V^2 + \Omega^2) 
[1-(\hat{\mathbf{s}} \cdot \hat{\mathbf{z}})^2]\, .
\end{equation}
This energy dissipation through the spin dynamics and the work rate done by the emergent electric field on the superflow are the same since
\begin{equation}
Q v \Omega = \frac{\alpha}{1 + \alpha^2} v^2
\end{equation}
and
\begin{equation}
\alpha (V^2 + \Omega^2) = \frac{\alpha}{1 + \alpha^2} v^2 \, .
\end{equation}

\subsection{$4\pi$ phase slips from a domain-wall dynamics}

Due to the U(1)$_{\phi+s}$ order parameter redundancy, the energy dissipation from the damping-induced precession in the domain-wall motion can be regarded as an equivalent of $4 \pi$ phase slips. As shown in Fig.~\ref{fig:fig2}, $\hat{\mathbf{u}}$ and $\hat{\mathbf{v}}$  rotate around $\hat{\mathbf{s}}$ by $\pm 2 \pi$ by adiabatically following the dynamics of the local spin direction $\hat{\mathbf{s}}$ in one cycle of precession. To see this, note that if we adopt the condition $\hat{\mathbf{v}} \cdot \hat{\mathbf{z}}=0$ in defining $\hat{\mathbf{v}}$, Eq.~\eqref{EQ:dwGen} will give us $\hat{\mathbf{v}} = -\hat{\mathbf{x}}\sin \Omega t + \hat{\mathbf{y}}\cos \Omega t$. But given the U(1)$_{\phi+s}$ redundancy, this is equivalent to the  $\pm 2 \pi$ phase twist on the left and the right end, respectively. 

The voltage arising from this precession can be understood either as arising from the emergent electric field, or equivalently, arising from the constant rate of $4\pi$ phase slips. When $\varphi$ increases at the rate $\dot{\varphi} = \Omega$, the emergent scalar potential at the two ends of the wire, $x = \pm \infty$, is given by
\begin{equation}
\tilde{V}_e = - \hat{\mathbf{s}} \cdot (\hat{\mathbf{u}} \times \partial_t \hat{\mathbf{u}}) = \begin{cases} - Q \Omega & \text{ at } x = -\infty \, , \\ Q \Omega & \text{ at } x = \infty \, , \end{cases}
\label{eq:Ve}
\end{equation}
which is exactly the Josephson voltage for the $4Q\pi$ phase slip occurring at the rate of $\Omega/2\pi$. Then, to maintain the finite superflow, there must be work given by 
\begin{equation}
W = v [\tilde{V}_{e} (x = \infty) - \tilde{V}_{e}(x = -\infty)] = 2 Q v \Omega \, ,
\end{equation}
by external reservoirs on the system, matching the work [Eq.~(\ref{eq:W})]. Figure~\ref{fig:fig2}(a) shows the time evolution of the triad \{$\hat{\mathbf{s}}, \hat{\mathbf{u}}, \hat{\mathbf{v}}$\} associated with the domain-wall that both moves and precesses. Note that $\hat{\mathbf{u}}$ (green arrow) at the left end ($x \rightarrow -\infty$) and the right end ($x \rightarrow \infty$) rotates clockwise and counterclockwise, respectively, about $\hat{\mathbf{s}}$ (red arrow), engendering the nonsingular phase slips across the $x$ direction. In a nutshell, the domain-wall angular dynamics produces the nonsingular $4 \pi$ phase slips that give rise to a finite voltage difference between the two ends of the superconducting wire, which constitutes our first main result.

There is a topological reason why one precession of a magnetic domain wall induces a $4 \pi$ phase slip, 
which can be derived from the Poincar\'e-Hopf theorem or Poincar\'e-Brower theorem~\cite{KamienRMP2002}. 
For concrete discussion on this, we will consider in the following 
the case with $Q = 1$ as shown in Fig.~\ref{fig:fig2}(b) and (c). 
At a given time, the instantaneous configuration of $\hat{\mathbf{s}}$ and $\hat{\mathbf{u}}$ can be mapped onto a line connecting the north pole [$\hat{\mathbf{s}} (x \rightarrow \infty)]$ and the south pole [$\hat{\mathbf{s}} (x \rightarrow - \infty)$] on the unit sphere by identifying $\hat{\mathbf{s}}$ with the surface normal as shown in Fig.~\ref{fig:fig2}(b). When we consider the collection of the configuration of \{$\hat{\mathbf{s}}, \hat{\mathbf{u}}$\} onto the unit sphere during one complete precession of the domain wall, i.e., $\varphi_0 \rightarrow \varphi_0 + 2 \pi$, $\hat{\mathbf{u}}$ can be regarded as a unit tangent vector field on the sphere since it is perpendicular to $\hat{\mathbf{s}}$, i.e., the surface normal as shown in Fig.~\ref{fig:fig2}(c). Here, note that $\hat{\mathbf{u}}$ is not uniquely determined at the north pole and the south pole, which is consistent with the well-known topological property of the sphere that the unit tangent vector field cannot be defined without a singularity on it. Instead of being uniquely determined, $\hat{\mathbf{u}}$ rotates by $2 \pi$ around the north pole and rotates by $- 2 \pi$ around the south pole as the domain wall completes one cycle of precession, which gives rise to a $4 \pi$ phase slip across the wire as discussed above [Eq.~(\ref{eq:Ve})]. This can be understood by applying the Poincar\'e-Hopf theorem to the unit tangent vector field $\hat{\mathbf{u}}$ on the sphere. The Euler number of the unit sphere is 2, meaning that the sum of the indices of the isolated singularities of the unit tangent vector field on the sphere must be 2. In our case, the indices of the north pole and the south pole associated with $\hat{\mathbf{u}}$ are both 1, adding up to 2, agreeing with the Euler number of the unit sphere.

\section{Meron}
\label{sec:meron}

\begin{figure}
\includegraphics[width=0.9\columnwidth]{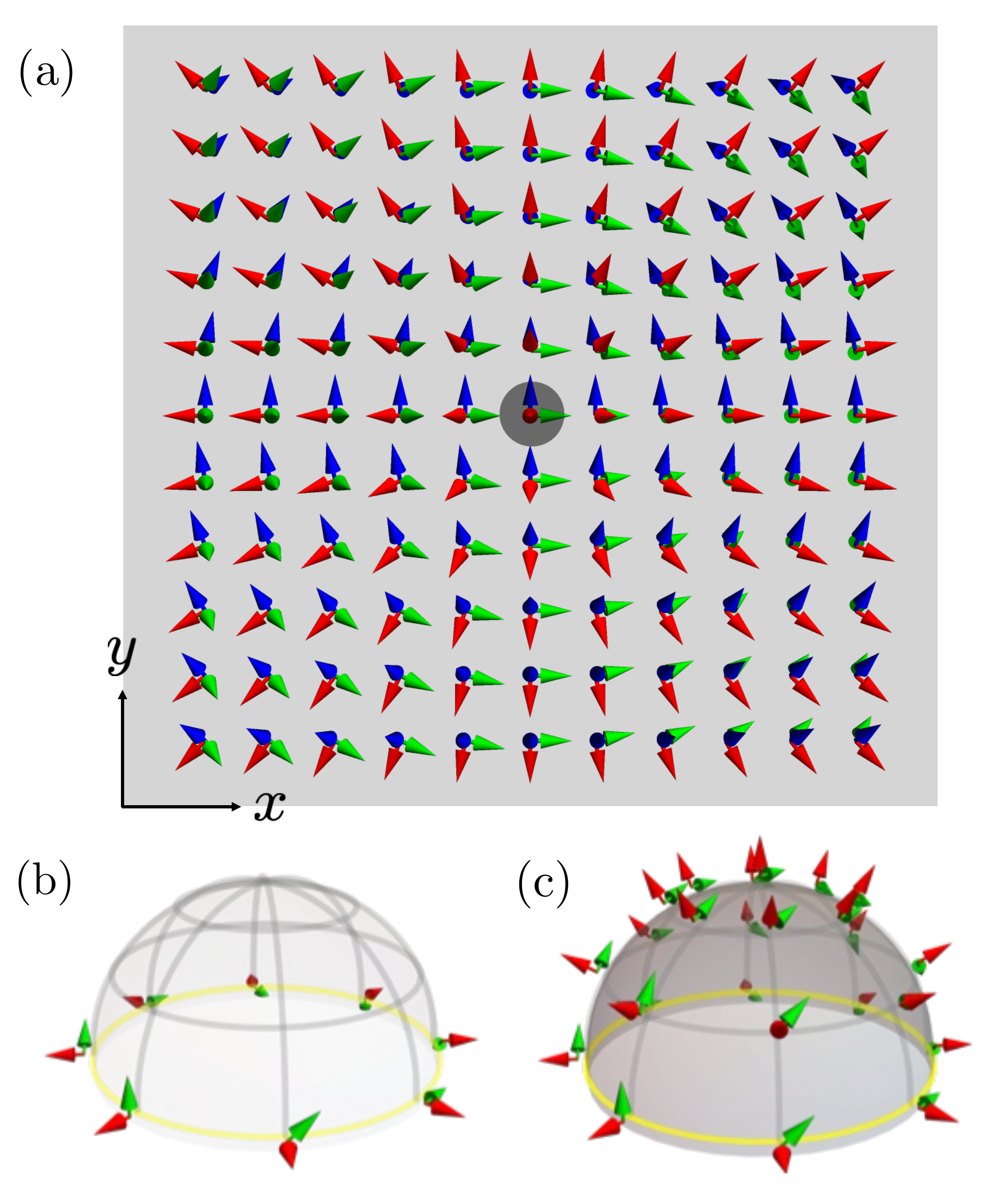}
\caption{(a) An illustration of the meron with polarity $p = 1$ and vorticity $n = 1$, which is a nonsingular topological defect of ferromagnetic superconductors with easy-plane spin anisotropy. The red, the green, and the blue arrows represent $\hat{\mathbf{s}}, \hat{\mathbf{u}}$, and $\hat{\mathbf{v}}$, respectively. The 
meron core is denoted by the gray dot. The local spin direction $\hat{\mathbf{s}}$ rotates by $2 \pi$ counterclockwise about the $z$ axis when we follow the infinitely distant trajectory encircling the meron 
core counterclockwise (and thus the vorticity $n = 1$). Note that $\hat{\mathbf{u}}$ and $\hat{\mathbf{v}}$ also change spatially to keep their orthonormality to $\hat{\mathbf{s}}$; they rotate by $2 \pi$ clockwise about the local spin direction $\hat{\mathbf{s}}$ when we enclose the meron center counterclockwise. (b) Mapping of the configuration of $\hat{\mathbf{s}}$ (red arrows) and $\hat{\mathbf{u}}$ (green arrows) along the 
infinitely distant circle in (a) onto the 
equator with the surface normal identified with $\hat{\mathbf{s}}$. (c) Mapping of the configuration of $\hat{\mathbf{s}}$ and $\hat{\mathbf{u}}$ of the entire system onto the northern hemisphere with $\hat{\mathbf{s}}$ identified with the surface normal. Note that $\hat{\mathbf{u}}$ is a well-defined unit tangent vector field 
on the northern hemisphere 
without any singularity. Since the Euler number of the hemisphere is 1, if the unit tangent vector field defined on the northern hemisphere has no singularity, it should rotate around the surface normal by $2 \pi$ along the equator according to the Poincar\'e-Hopf theorem~\cite{KamienRMP2002}, which is exactly what $\hat{\mathbf{u}}$ (unit tangent vector) does around $\hat{\mathbf{s}}$ (surface normal).}
\label{fig:v}
\end{figure}

We now consider ferromagnetic superconductors with easy-plane spin anisotropy ($\nu = \sgn(D) = -1$). For easy-plane spin systems, a meron, with its one-half skyrmion charge, is a generically stable topological defect~\cite{GrossNPB1978, EzawaPRB2011, LinPRB2015, YuNature2018}. Therefore, we consider the rotational Cooper pair velocity in absence of the applied magnetic field {\it i.e.} $\boldsymbol{\nabla} \times {\bf v} = -{\bf b}/\tilde{m}$ with $h=0$. See Fig.~\ref{fig:v} for the schematic illustration of a meron.

\subsection{Static solution}

Analogous to the case of the domain wall in the previous section, a straightforward construction of the meron motion solution here in the background of the constant and uniform background superflow is possible from the static meron solution in absence of any background superflow for the incompressible limit, $\eta \rightarrow \infty$~\cite{LamacraftPRA2008, BarnettPRB2009}. 
The static solution can be obtained from the following two equations; 
\begin{eqnarray}
\tilde{m} (\boldsymbol{\nabla} \times \mathbf{v}) &=& - \mathbf{b} \, , \nonumber\\
(\mathbf{v} \cdot \boldsymbol{\nabla}) \hat{\mathbf{s}} &=& \partial_i [(\hat{\mathbf{s}} \times \partial_i \hat{\mathbf{s}})] - (\hat{\mathbf{s}} \cdot \hat{\mathbf{z}}) \hat{\mathbf{s}} \times \hat{\mathbf{z}} \, .
\label{EQ:meronStaticEq}
\end{eqnarray}
It is important to note here that while we are dealing with a static configuration we still have ${\bf v} \neq 0$ due to the intrinsic emergent magnetic flux of the meron. In addition, this solution would possess the axial symmetry, {\it i.e.} 
\begin{equation}
\hat{\mathbf{s}}_0 = (\sin \theta \cos \varphi, \sin \theta \sin \varphi, \cos \theta)\, ,
\label{EQ:meronStatic}
\end{equation}
with $\theta = \theta(r)$ and $\varphi = n\chi + \Phi$ where $(r, \chi)$ are polar coordinates for the two-dimensional system, and follow the universal boundary conditions for merons are given by 
$$
\theta(r=0) = (1 - p) \frac{\pi}{2}\, , \quad \theta(r \to \infty) = \frac{\pi}{2}\, ;
$$ 
$p = \pm 1$ here is the polarity, which is the $z$-component of the local spin direction $\hat{\mathbf{s}}$ at the meron center, and $n \in \mathbb{Z}$ the vorticity, which counts how many times $\hat{\mathbf{s}}$ winds in the easy plane along the closed trajectory encircling the meron center. For our purpose, obtaining the differential equation for $\theta(r)$ is sufficient for showing Eq.~\eqref{EQ:meronStatic} to be the solution of Eq.~\eqref{EQ:meronStaticEq}. We start by noting that the emergent magnetic field is aligned entirely along the $z$-axis and 
\begin{equation}
b_z = - \hat{\mathbf{s}}_0 \cdot (\partial_x \hat{\mathbf{s}}_0 \times \partial_y \hat{\mathbf{s}}_0) = - \frac{n \sin \theta}{r} \frac{d \theta}{d r} \, 
\end{equation}
is function only for $\theta$, which gives us the well-known result 
\begin{equation}
\int dx dy b_z = \int r d\chi dr \left( - \frac{n \sin \theta}{r} \frac{d \theta}{d r} \right) =  - 2 \pi p n \, 
\label{EQ:fluxQuant}
\end{equation}
for the total emergent magnetic flux. With this emergent magnetic field, the first equation of Eq.~\eqref{EQ:meronStaticEq} requires the circulating velocity $\mathbf{v} = \hat{\boldsymbol{\varphi}} v(r) $ around the meron with
$$
\frac{1}{r} \frac{d (r v)}{d r} = \frac{n \sin \theta}{\tilde{m} r} \frac{d \theta}{d r} \, .
$$
Inserting this relation into the second equation of Eq.~\eqref{EQ:meronStaticEq} gives us
$$ 
\sin \theta (1 - \cos \theta) \frac{1}{\tilde{m}r^2}=\frac{1}{r} \frac{d}{d r} \left( r \frac{d \theta}{dr} \right) + \cos \theta \sin \theta \left( 1 - \frac{1}{r^2} \right)   \, ,
$$
from which $\theta(r)$ can be obtained numerically.

\begin{figure*}
\includegraphics[width=1.8\columnwidth]{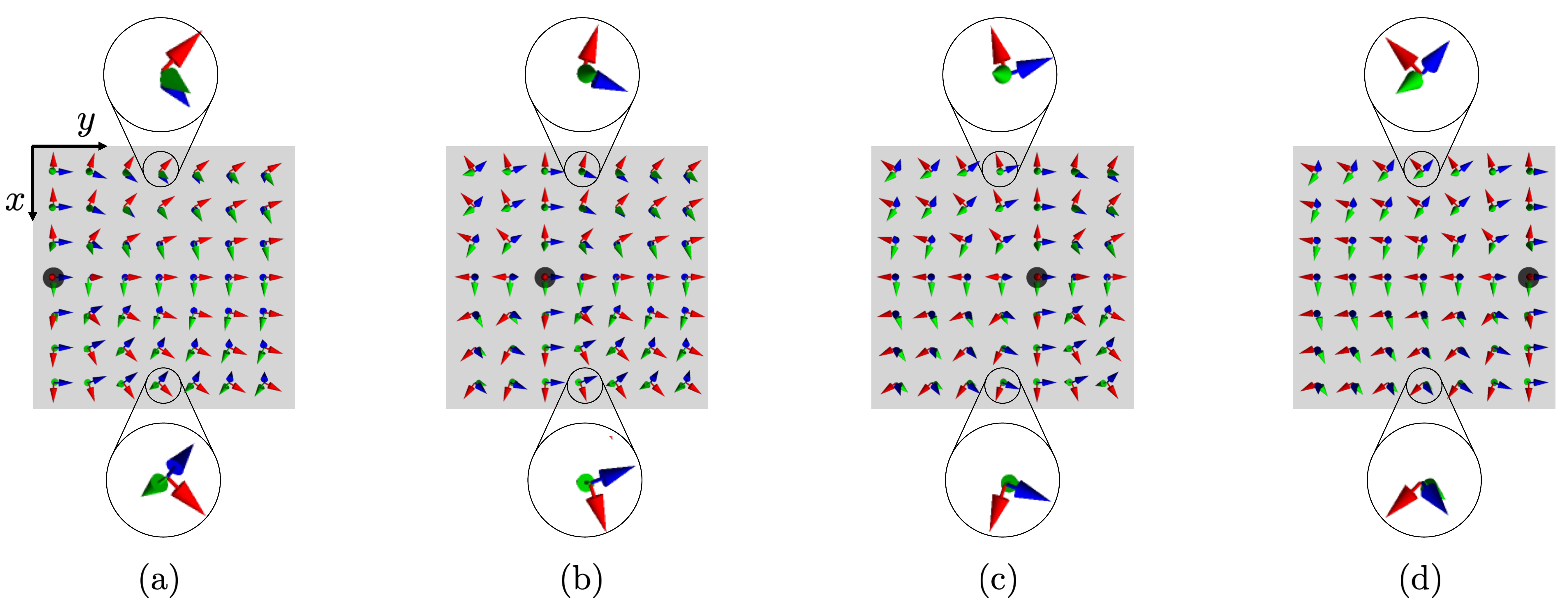}
\caption{Snapshots of a meron moving in the $y$ direction in the increasing time from (a) to (d), where the meron core is depicted by a gray dot. At $x \to - \infty$, $\hat{\mathbf{u}}$ (green arrow) rotates about $\hat{\mathbf{s}}$ (red arrow) counterclockwise, whereas at $x \to \infty$, $\hat{\mathbf{u}}$ rotates about $\hat{\mathbf{s}}$ clockwise, inducing phase slips and thereby generating a finite voltage in the $x$ direction.}
\label{fig:fig4}
\end{figure*}

To find explicit expressions for $\hat{\mathbf{u}}$ and $\hat{\mathbf{v}}$, note that the following three unit vectors form an orthonormal triad:
\begin{eqnarray*}
\hat{\mathbf{s}}_0 = \hat{\mathbf{e}}_r &\equiv& (\sin \theta \cos \varphi, \sin \theta \sin \varphi, \cos \theta) \, , \nonumber\\
\hat{\mathbf{e}}_\theta &\equiv& (\cos \theta \cos \varphi, \cos \theta \sin \varphi, - \sin \theta) \, , \nonumber\\
\hat{\mathbf{e}}_\varphi &\equiv& (- \sin \varphi, \cos \varphi, 0) \, ,
\end{eqnarray*}
which gives us
\begin{eqnarray}
\hat{\mathbf{u}}_0 (r, \chi) = \cos \varphi(\chi) \, \hat{\mathbf{e}}_\theta(r, \chi)- \sin \varphi(\chi) \, \hat{\mathbf{e}}_\varphi (r, \chi) \, , \label{eq:u} \\
\hat{\mathbf{v}}_0 (r, \chi) = \sin \varphi(\chi) \, \hat{\mathbf{e}}_\theta(r, \chi) + \cos \varphi(\chi) \, \hat{\mathbf{e}}_\varphi (r, \chi) \, . \label{eq:v}
\end{eqnarray}

The local configuration of the triad $(\hat{\mathbf{s}}_0, \hat{\mathbf{u}}_0, \hat{\mathbf{v}}_0)$ for a meron with $p = 1$ and $n = 1$ is shown in Fig.~\ref{fig:v}. Note that it is nonsingular, differing from a conventional vortex of a $s$-wave superconductor~\cite{HalperinJLTP1979}. 
An analogous nonsingular 
topological defects that give rise to $4 \pi$ nonsingular phase slips has been discussed by \textcite{AndersonPRL1977} 
and has been termed the skyrmion solution 
in the more recent literature~\cite{LiPRB2009, BarnettPRB2009}. By contrast, our solution \{$\hat{\mathbf{s}}_0, \hat{\mathbf{u}}_0, \hat{\mathbf{v}}_0$\} [Eqs.~(\ref{EQ:meronStatic},\ref{eq:u},\ref{eq:v})] represents an explicit solution for the nonsingular topological defect in the easy-plane case that gives rise to $2 \pi$ nonsingular phase slip, as can be seen from Eqs~\eqref{EQ:meronStaticEq} and \eqref{EQ:fluxQuant}: 
it harbors the emergent magnetic flux $- 2 \pi p n$ and thus its motion gives rise to the emergent electric field, {\it i.e.}, phase slips perpendicular to its motion.

The non-trivial emergent gauge field $a_i$, and thus the quantized non-zero emergent magnetic flux, of our nonsingular topological defect 
can be understood from the rotation of $\hat{\mathbf{u}}$ around $\hat{\mathbf{s}}$ as stated in Eq.~(\ref{EQ:eGauge}). There exists a topological constraint dictating that a 
meron texture of $\hat{\mathbf{s}}$ should trap a quantized non-zero emergent magnetic flux, 
which can be understood by invoking the Poincar\'e-Hopf theorem~\cite{KamienRMP2002, Volovik2003} as follows. For the given 
meron configuration with $p = 1$, let us consider the mapping of two unit vectors $\hat{\mathbf{s}}$ and $\hat{\mathbf{u}}$ onto the northern hemisphere such that $\hat{\mathbf{s}}$ is identified with the surface normal. Then, $\hat{\mathbf{u}}$ becomes a unit tangent vector field on the hemisphere since it is perpendicular to $\hat{\mathbf{s}}$, i.e., the surface normal as shown in Fig.~\ref{fig:v}(c). The Euler number of the hemisphere is 1, and thus the Poincar\'e-Hopf theorem dictates that, if the unit tangent vector field is nonsingular on the northern hemisphere, it should rotate exactly one time about the surface normal while traversing the equator, which is exactly what $\hat{\mathbf{u}}$ does around $\hat{\mathbf{s}}$ in Fig.~\ref{fig:v}(b). Therefore, the one-time rotation of $\hat{\mathbf{u}}$ around $\hat{\mathbf{s}}$ along the closed loop that contains, but is also infinitely far from, the meron core as shown in Fig.~\ref{fig:v}(a), which gives rise to the quantized emergent magnetic flux, can be regarded as the physical manifestation of the topological constraint that should be satisfied by the nonsingular unit tangent vector field defined on the hemisphere.

\subsection{Dynamics with a background superflow}

Analogous to the domain wall motion, it is straightforward to work out the spin equation of motion for the meron motion driven by a uniform constant background superflow $\mathbf{v}_0$  if we assume the meron motion to be rigid, {\it i.e.} $\hat{\mathbf{s}}(\mathbf{r}, t) = \hat{\mathbf{s}}_0 (\mathbf{r} - \mathbf{V} t)$ (the same holds for $\hat{\mathbf{u}}$ and $\hat{\mathbf{v}}$). The Cooper pair velocity would then be given by ${\bf v} = {\bf v}_m + {\bf v}_0$, where $\mathbf{v}_m$ is the Cooper pair velocity around a static meron. We can therefore employ the collective coordinate approach to describe the dynamics of a meron, and use the fact that $\hat{\mathbf{s}}_0 (\mathbf{r} - \mathbf{V} t)$ should satisfy the original spin equation of motion of Eq.~\eqref{EQ:GDamp} with, as we are in the incompressible limit, the constant $\tilde{\rho}$ while $\hat{\mathbf{s}}_0 (\mathbf{r})$ also satisfies the equation for the static meron of Eq.~\eqref{EQ:meronStaticEq}. Subtraction between the two equations, together with $\partial_t \hat{\mathbf{s}}_0 (\mathbf{r} - \mathbf{V} t) = -{\bf V} \cdot \boldsymbol{\nabla} \hat{\mathbf{s}}_0 (\mathbf{r} - \mathbf{V} t)$ gives us
$$
- (1+\alpha \hat{\mathbf{s}}_0 \times) {\bf V} \cdot \boldsymbol{\nabla} \hat{\mathbf{s}}_0 = -{\bf v}_0 \cdot \boldsymbol{\nabla} \hat{\mathbf{s}}_0\, .
$$
Taking the scalar product with $\mathbf{s}_0 \times \partial_i \mathbf{s}_0$ on both sides gives us 
\begin{equation}
\begin{split}
& - V_j [\mathbf{s}_0 \cdot (\partial_i \hat{\mathbf{s}}_0 \times \partial_j \hat{\mathbf{s}}_0)] - \alpha V_j (\partial_i \hat{\mathbf{s}}_0 \cdot \partial_j \hat{\mathbf{s}}_0) \\
= & - v_{0,j} [\mathbf{s}_0 \cdot (\partial_i \hat{\mathbf{s}}_0 \times \partial_j \hat{\mathbf{s}}_0)] \, .
\end{split}
\label{eq:meron}
\end{equation}
Strictly speaking, this result shows that whereas we have an exact rigid motion solution with ${\bf V} = {\bf v}_0$ in absence of damping, no rigid motion solution can be exact in presence of damping. Yet to the zeroth order in $v_0$ and also in the spirit of collective coordinate, we can average out the effect of the spin texture, {\it i.e.} defining 
$$
G_{ij} = \int dx dy \mathbf{s}_0 \cdot (\partial_i \hat{\mathbf{s}}_0 \times \partial_j \hat{\mathbf{s}}_0) \, , \quad D_{ij} =  \int dx dy \partial_i \hat{\mathbf{s}}_0 \cdot \partial_j \hat{\mathbf{s}}_0 \, ,
$$
known respectively as the gyrotropic coefficients and the dissipation coefficients~\cite{ThielePRL1973, IvanovPRL1994, TretiakovPRL2008} ($D_{ij} = D \delta_{ij}$ and $G_{xy} = - G_{yx} \equiv G = 2 \pi p n$ for a meron). By integrating Eq.~(\ref{eq:meron}), we have
\begin{equation}
\left( \alpha D + G \hat{\mathbf{z}} \times \right) \mathbf{V} = G \hat{\mathbf{z}} \times \mathbf{v}_0 \, ,
\end{equation}
which yields the following solution for the velocity $\mathbf{V}$:
\begin{equation}
\mathbf{V} = \frac{G}{G^2 + \alpha^2 D^2} \left(G + \alpha D \hat{\mathbf{z}} \times \right) \mathbf{v}_0 \, .
\end{equation}
To consider a concrete example, we will hereafter restrict the discussion to the case where the background superflow flows in the $x$ direction: $\mathbf{v}_{0} = (v_0, 0)$. In this case, we have
\begin{equation}
\begin{pmatrix} V_x \\ V_y \end{pmatrix} = \frac{G}{G^2 +\alpha^2 D^2} \begin{pmatrix} G v_0 \\ - \alpha D v_0 \end{pmatrix} \, .
\end{equation}
Note that the presence of damping gives rise to the component of the meron velocity transverse to the uniform superflow in proportion to the skyrmion number $G/2\pi$ of the meron, $V_y = - \alpha[GD / (G^2 +\alpha^2 D^2)] v_0$, exhibiting the so-called skyrmion Hall effect~\cite{ZangPRL2011, Everschor-SitteJAP2014, LitzuisNP2016, JiangNP2017}. This transverse motion of the meron with respect to the superflow gives rise to the finite voltage in the direction of the superflow via the $2 \pi$ phase slips, which we turn our attention now.

\subsection{$2\pi$ phase slips from a meron motion}

Again analogous to the domain wall motion, the U(1)$_{\phi+s}$ order parameter redundancy allows the energy dissipation due to the meron motion as equivalent to the $2\pi$ phase slips. This can be seen from the emergent electric field
\begin{eqnarray}
{\bf e} &=& - \hat{\mathbf{s}} \cdot (\boldsymbol{\nabla} \hat{\mathbf{s}} \times \partial_t \hat{\mathbf{s}}) \,  \nonumber\\
&=& \hat{\mathbf{s}}_0 \cdot (\boldsymbol{\nabla} \hat{\mathbf{s}}_0 \times {\bf V} \cdot \boldsymbol{\nabla} \hat{\mathbf{s}}_0) = - {\bf V} \times {\bf b} \, ,
\end{eqnarray}
which is in the same form as the Josephson electric field arising from the vortex motion~\cite{HalperinJLTP1979}. The same can naturally be said about the input power density required for driving the uniform constant background superflow
\begin{eqnarray*}
- \mathbf{v}_0 \cdot \mathbf{e} &=& v_{0,x} \hat{\mathbf{s}} \cdot (\partial_x \hat{\mathbf{s}} \times \partial_t \hat{\mathbf{s}}) \, , \nonumber\\
&=& - v_{0,x} V_y \hat{\mathbf{s}} \cdot (\partial_x \hat{\mathbf{s}} \times \partial_y \hat{\mathbf{s}}) \, .
\end{eqnarray*}
It can be checked explicitly that the total energy rate for driving the superflow
\begin{equation*}
- \int dx dy \mathbf{v}_0 \cdot \hat{\mathbf{e}} = - v_{0,x} V_y G = \alpha\frac{G^2 D}{G^2 + \alpha^2 D^2} v_0^2 \, ,
\end{equation*}
is equal to the energy dissipation rate
\begin{eqnarray*}
\int dxdy \alpha (\partial_t \hat{\mathbf{s}})^2 &=&\int dxdy \alpha V_i V_j (\partial_i \hat{\mathbf{s}}_0 \cdot \partial_j \hat{\mathbf{s}}_0)\nonumber\\ 
&=& \alpha D \mathbf{V}^2 = \alpha\frac{G^2 D}{G^2 +\alpha^2  D^2} v_0^2 \, .
\end{eqnarray*}

This energy must come from external reservoirs through the boundary of the system, meaning that there should be a development of a finite voltage across the system in the $x$ direction. This can be explicitly seen from 
\begin{equation}
\tilde{V}_e = - \hat{\mathbf{s}} \cdot (\hat{\mathbf{u}} \times \partial_t \hat{\mathbf{u}}) = (1 - \cos \theta) \partial_t \varphi \, .
\end{equation}
To see how a finite voltage is generated by the motion of a vortex, let us assume that a vortex moves in the $y$ direction, $\mathbf{V} = V_y \hat{\mathbf{y}}$. Then for a given point at large $x$, 
\begin{equation}
\int_{-\infty}^{\infty} dt \hat{\mathbf{s}} \cdot (\hat{\mathbf{u}} \times \partial_t \hat{\mathbf{u}}) = - \int_{-\infty}^{\infty} dt \partial_t \varphi = n \pi \, , \quad \text{for } x \to +\infty \, .
\end{equation}
The vector $\hat{\mathbf{u}}(x \rightarrow \infty)$ rotates by $n \pi$ around the $\hat{\mathbf{s}}$. Also,
\begin{equation}
\int_{-\infty}^{\infty} dt \hat{\mathbf{s}} \cdot (\hat{\mathbf{u}} \times \partial_t \hat{\mathbf{u}}) = - \int_{-\infty}^{\infty} dt \partial_t \varphi = - n \pi \, , \quad \text{for } x \to -\infty \, .
\end{equation}
The vector $\hat{\mathbf{u}}(x \rightarrow - \infty)$ rotates by $- n \pi$ around the $\hat{\mathbf{s}}$. This indicates that the motion of a meron in the $y$ direction induces a nonsingular $2 \pi$ phase slips across the $x$ direction. To keep the superflow in the $x$ direction constant, we need to counteract the effect of these phase slips. The corresponding work on the system is dissipated to external baths such as quasiparticles or phonons through the Gilbert damping. Figure~\ref{fig:fig4} shows the schematic illustration of the process. As the vortex moves in the positive $y$ direction from Fig.~\ref{fig:fig4}(a) to Fig.~\ref{fig:fig4}(d), $\hat{\mathbf{u}}$ at $x \rightarrow \infty$ and $x \rightarrow - \infty$ rotates about the local spin direction $\hat{\mathbf{s}}$ counterclockwise and clockwise, respectively, producing phase slips in the $x$ direction. This is our second main result: The dynamics of a meron engenders the nonsingular $2 \pi$ phase slips perpendicular to its motion through the generation of the emergent electric field, showcasing the intertwined dynamics of spin and charge degrees of freedom of ferromagnetic superconductors.

\section{Discussion}
\label{sec:discussion}

Within the phenomenological framework for the dynamics of the order parameter of ferromagnetic superconductors, we have shown that the current-induced dynamics of magnetic defects in the presence of the spin dissipation, i.e., the Gilbert damping, give rise to nonsingular phase slips via the emergent electromagnetic fields. The input power, which is the product of the applied current that drives the magnetic defects and the voltage generated by the phase slip, is shown to be equivalent to the dissipated energy in the form the Gilbert damping into the baths of quasiparticles or phonons. Our work on the dynamics of magnetic defects showcases the intrinsic interplay of spin and charge dynamics of ferromagnetic superconductors. 

A few remarks are on order about the limitations of our work. First, we did not include the effects of the non-adiabatic spin-transfer torque by a supercurrent on the dynamics of magnetic defects, which is expected to be present on general grounds whenever the Gilbert damping is present~\cite{ZhangPRL2004, TataraPR2008, TserkovnyakJMMM2008, Takashima2017}. While we believe that the inclusion of the non-adiabatic spin-transfer torque in our model would not qualitatively change the relations that we have found between the dynamics of magnetic defects and nonsingular phase slips, it will certainly enrich the physics of the interplay of spin and charge dynamics in ferromagnetic superconductors. Secondly, in this work, a ferromagnetic superconductor is assumed to be a fully spin-polarized triplet superconductor as in Ref.~\cite{CornfeldPRR2021}, by leaving the generalization to a partially spin-polarized case as future work. Thirdly, our analysis shows that the assumption of rigid motion is not exact for merons in presence of damping. The coupling between current and magnons bound to meron cores may be a relevant topic for future study. Lastly, the dynamics of magnetic defects has been discussed in the incompressible limit, where the dynamics of the order-parameter amplitude is frozen. Releasing this assumption would allow us to study the interplay of spin dynamics and longitudinal order-parameter dynamics, which is beyond the scope of the current work.

\begin{acknowledgments}
We thank Mike Stone, Grigori Volovik, Daniel Agterberg and Jim Sauls for useful discussions. S.K.K. was supported by Brain Pool Plus Program through the National Research Foundation of Korea funded by the Ministry of Science and ICT (NRF-2020H1D3A2A03099291) and by the National Research Foundation of Korea funded by the Korea Government via the SRC Center for Quantum Coherence in Condensed Matter (NRF-RS-2023-00207732). S.B.C. was supported by the National Research Foundation of Korea (NRF) grants funded by the Korea government (MSIT) (NRF-2023R1A2C1006144, NRF-2020R1A2C1007554, and NRF-2018R1A6A1A06024977).
\end{acknowledgments}

\newpage
\begin{appendix}
\begin{widetext}
\section{Details of the order parameter}
\label{app:op}

The multicomponent superconducting gap is given by
\begin{equation}
\hat{\Delta} = \begin{pmatrix} \Delta_{\uparrow \uparrow} & \Delta_{\uparrow \downarrow} \\ \Delta_{\downarrow \uparrow} & \Delta_{\downarrow \downarrow} \end{pmatrix} \equiv \begin{pmatrix} - \tilde{d}_x + i \tilde{d}_y & \tilde{d}_z \\ \tilde{d}_z & \tilde{d}_x + i \tilde{d}_y \end{pmatrix} = i (\mathbf{d} \cdot \boldsymbol{\sigma}) \sigma_y \, .
\end{equation}

Then,
\begin{eqnarray}
\hat{\Delta} \hat{\Delta}^\dagger &=& \begin{pmatrix} \Delta_{\uparrow \uparrow} & \Delta_{\uparrow \downarrow} \\ \Delta_{\downarrow \uparrow} & \Delta_{\downarrow \downarrow} \end{pmatrix} \begin{pmatrix} \Delta_{\uparrow \uparrow}^* & \Delta_{\downarrow \uparrow}^* \\ \Delta_{\uparrow \downarrow}^* & \Delta_{\downarrow \downarrow}^* \end{pmatrix} \, \nonumber\\
&=& \begin{pmatrix} |\Delta_{\uparrow \uparrow}|^2 + |\Delta_{\uparrow \downarrow}|^2 & \Delta_{\uparrow \uparrow} \Delta_{\downarrow \uparrow}^* + \Delta_{\uparrow \downarrow} \Delta_{\downarrow \downarrow}^* \\ \Delta_{\downarrow \uparrow} \Delta_{\uparrow \uparrow}^* + \Delta_{\downarrow \downarrow} \Delta_{\uparrow \downarrow}^* & |\Delta_{\downarrow \downarrow}|^2 + |\Delta_{\downarrow \uparrow}|^2 \end{pmatrix} \, \nonumber\\
&=& \begin{pmatrix} - \tilde{d}_x + i \tilde{d}_y & \tilde{d}_z \\ \tilde{d}_z & \tilde{d}_x + i \tilde{d}_y \end{pmatrix} \begin{pmatrix} - \tilde{d}^*_x - i \tilde{d}^*_y & \tilde{d}^*_z \\ \tilde{d}^*_z & \tilde{d}^*_x - i \tilde{d}^*_y \end{pmatrix} \, \nonumber\\
&=& \begin{pmatrix} |\tilde{d}_x|^2 + |\tilde{d}_y|^2 + |\tilde{d}_z|^2 + i (\tilde{d}_x \tilde{d}_y^* - \tilde{d}^*_x \tilde{d}_y) & (- \tilde{d}_x \tilde{d}_z^* + i \tilde{d}_y \tilde{d}_z^*) + \text{c.c.} \\ (- \tilde{d}_x^* - i \tilde{d}_y^*) \tilde{d}_z) + \text{c.c.} & |\tilde{d}_x|^2 + |\tilde{d}_y|^2 + |\tilde{d}_z|^2 - i (\tilde{d}_x \tilde{d}_y^* - \tilde{d}^*_x \tilde{d}_y) \end{pmatrix} \, \nonumber\\
&=& |\mathbf{d}|^2 \sigma_0 + i (\mathbf{d} \times \mathbf{d}^*) \cdot \boldsymbol{\sigma} \, ,
\end{eqnarray}
where $\mathbf{d} = (\tilde{d}_x, \tilde{d}_y, \tilde{d}_z)$. The total number density of the Cooper pairs is given by
\begin{equation}
|\Delta_{\uparrow \uparrow}|^2 + |\Delta_{\uparrow \downarrow}|^2+ |\Delta_{\downarrow \uparrow}|^2 + |\Delta_{\downarrow \downarrow}|^2 = \Tr [\hat{\Delta} \hat{\Delta}^\dagger] = 2 \mathbf{d} \cdot \mathbf{d}^* = 2 |\mathbf{d}|^2 \, .
\end{equation}

The expected spin angular momentum is given in units of $\hbar$ by
\begin{equation}
\Tr[ \hat{\Delta} \hat{\Delta}^\dagger \boldsymbol{\sigma} ] = \Tr[ i (\mathbf{d} \times \mathbf{d}^*) \cdot \boldsymbol{\sigma} \boldsymbol{\sigma} ] = 2 i \mathbf{d} \times \mathbf{d}^* \, .
\end{equation}

For a fully spin-polarized triplet superconductor, we can use
\begin{equation}
\mathbf{d} = d e^{i \phi} (\hat{\mathbf{u}} + i \hat{\mathbf{v}}) \, ,
\end{equation}
with $\hat{\mathbf{u}} \perp \hat{\mathbf{v}}$. Then, we have $\mathbf{d} \cdot \mathbf{d}^* = 2 d^2$. Therefore, $d = \sqrt{\rho}/2$. The spin polarization is then given by
\begin{eqnarray}
\mathbf{s} &=& 2 i \mathbf{d} \times \mathbf{d}^* = 4 d^2 (\hat{\mathbf{u}} \times \hat{\mathbf{v}}) \, \nonumber\\
&\equiv& 4 d^2 \hat{\mathbf{s}} = \rho \hat{\mathbf{s}} \, .
\end{eqnarray}

To see if the result makes sense, let us consider $\hat{\mathbf{s}} = \hat{\mathbf{z}}$ with $\hat{\mathbf{u}} = \hat{\mathbf{x}}$ and $\hat{\mathbf{v}} = \hat{\mathbf{y}}$. Then, $\mathbf{d} = d e^{i \phi} (\hat{\mathbf{x}} + i \hat{\mathbf{y}})$. Then, $\tilde{d}_x = d e^{i \phi}$ and $\tilde{d}_y = d e^{i \phi} i$. Then, $\Delta_{\uparrow \uparrow} = - 2 d, \Delta_{\downarrow \downarrow} = \Delta_{\uparrow \downarrow} = \Delta_{\downarrow \uparrow} = 0$. The condensate density (of the Cooper pairs) is given by $\rho = |\Delta_{\uparrow \uparrow}|^2 = 4 d^2$ and the spin density is given by $\mathbf{s} = 4 d^2 \hat{\mathbf{s}} = \rho \hat{\mathbf{s}}$.

\section{Kinetic term of the Lagrangian}
\label{app:kin}

The kinetic term of the Lagrangian density is given by
\begin{eqnarray}
\mathcal{L}_K &=& 2 i \hbar \mathbf{d}^* \cdot \partial_t \mathbf{d} \, \nonumber\\
&=& \frac{i \hbar}{2} \sqrt{\rho} e^{- i \phi} (\hat{\mathbf{u}} - i \hat{\mathbf{v}}) \cdot \partial_t \left[ \sqrt{\rho} e^{i \phi} (\hat{\mathbf{u}} + i \hat{\mathbf{v}}) \right] \, \nonumber\\
&=& \frac{i \hbar}{2} \sqrt{\rho} e^{- i \phi} (\hat{\mathbf{u}} - i \hat{\mathbf{v}}) \cdot \left[ \frac{\partial_t \rho}{2 \sqrt{\rho}} e^{i \phi} (\hat{\mathbf{u}} + i \hat{\mathbf{v}}) + (i \partial_t \phi) \sqrt{\rho} e^{i \phi} (\hat{\mathbf{u}} + i \hat{\mathbf{v}}) + \sqrt{\rho} e^{i \phi} (\partial_t \hat{\mathbf{u}} + i \partial_t \hat{\mathbf{v}}) \right] \, \nonumber\\
&=& \frac{i \hbar}{2} \sqrt{\rho} (\hat{\mathbf{u}} - i \hat{\mathbf{v}}) \cdot \left[ \frac{\partial_t \rho}{2 \sqrt{\rho}} (\hat{\mathbf{u}} + i \hat{\mathbf{v}}) + (i \partial_t \phi) \sqrt{\rho} (\hat{\mathbf{u}} + i \hat{\mathbf{v}}) + \sqrt{\rho} (\partial_t \hat{\mathbf{u}} + i \partial_t \hat{\mathbf{v}}) \right] \, \nonumber\\
&=& \frac{i \hbar}{2} \left[ \frac{\partial_t \rho}{2} 2 + (i \partial_t \phi) \rho 2 + \rho (i \hat{\mathbf{u}} \cdot \partial_t \hat{\mathbf{v}} - i \hat{\mathbf{v}} \cdot \partial_t \hat{\mathbf{u}}) \right] \, \nonumber\\
&=& - \hbar \rho (\partial_t \phi + \hat{\mathbf{u}} \cdot \partial_t \hat{\mathbf{v}}) \, \nonumber\\
&=& - \hbar \rho [\partial_t \phi -\hat{\mathbf{s}} \cdot (\hat{\mathbf{u}} \times \partial_t \hat{\mathbf{u}})] \, .
\end{eqnarray}
The factor of $2$ in front is to ensure the commutation relation, $[ d^*_i (\mathbf{r}), d_j (\mathbf{r}') ] = 2 i \hbar \delta(\mathbf{r} - \mathbf{r}')$. Then, the total Lagrangian density is given by
\begin{equation}
\begin{split}
\mathcal{L} = & - \hbar \rho [\partial_t \phi -\hat{\mathbf{s}} \cdot (\hat{\mathbf{u}} \times \partial_t \hat{\mathbf{u}})] \\
& - \rho \left\{ \frac{A_c}{2} \left[ \partial_i \phi + \frac{q}{\hbar} A_i + \hat{\mathbf{s}} \cdot (\hat{\mathbf{u}} \times \partial_i \hat{\mathbf{u}}) \right]^2 + \frac{A_s}{2 } (\partial_i \hat{\mathbf{s}})^2 - \frac{D}{2} (\hat{\mathbf{s}} \cdot \hat{\mathbf{z}})^2 - H s_z + q V \right\} \\
& - \left\{ \frac{A_c}{16} (\boldsymbol{\nabla} \rho)^2 + \frac{U}{2} (\rho - \rho_0)^2 \right\} \, .
\end{split}
\end{equation}

From this, we can read the emergent scalar potential
\begin{equation}
V_e = - \frac{\hbar}{q} \hat{\mathbf{s}} \cdot (\hat{\mathbf{u}} \times \partial_t \hat{\mathbf{u}}) \, ,
\end{equation}
and the emergent vector potential
\begin{equation}
a_i = \frac{\hbar}{q} \hat{\mathbf{s}} \cdot (\hat{\mathbf{u}} \times \partial_i \hat{\mathbf{u}}) \, .
\end{equation}

\section{Equations of motion}
\label{app:eom}

The dynamics of the order parameter can be uniquely characterized by the dynamics of three variables, the condensate number density of the Cooper pairs $\rho$, the phase of the order parameter $\phi$, and the spin direction $\hat{\mathbf{s}}$.

First, the equation of motion for $\phi$ can be obtained by
\begin{eqnarray}
&& \frac{\delta L}{\delta \rho} = 0 \, ,\nonumber\\
&\Rightarrow& - \hbar [\partial_t \phi -\hat{\mathbf{s}} \cdot (\hat{\mathbf{u}} \times \partial_t \hat{\mathbf{u}})] = \frac{\delta F}{\delta \rho} \, ,\nonumber\\
&\Rightarrow& - \hbar [\partial_t \phi -\hat{\mathbf{s}} \cdot (\hat{\mathbf{u}} \times \partial_t \hat{\mathbf{u}})] = \left\{ \frac{A_c}{2} \left[ \partial_i \phi - \frac{q}{\hbar} A_i - \hat{\mathbf{s}} \cdot (\hat{\mathbf{u}} \times \partial_i \hat{\mathbf{u}}) \right]^2 + \frac{A_s}{2 } (\partial_i \hat{\mathbf{s}})^2 - \frac{D}{2} (\hat{\mathbf{s}} \cdot \hat{\mathbf{z}})^2 - H s_z + q V \right\} - \frac{A_c}{8} \nabla^2 \rho + U (\rho - \rho_0) \, .\nonumber\\
\end{eqnarray}

Second, the equation of motion for $\rho$ can be obtained by
\begin{eqnarray}
&& \frac{d}{dt} \left( \frac{\delta L}{\delta \dot{\phi}} \right) - \frac{\delta L}{\delta \phi} = 0 \, ,\nonumber\\
&\Rightarrow& - \hbar \dot{\rho} = - \frac{\delta F}{\delta \phi} \, ,\nonumber\\
&\Rightarrow& - \hbar \dot{\rho} = \partial_i \left\{ \rho A_c \left[ \partial_i \phi - \frac{q}{\hbar} A_i - \hat{\mathbf{s}} \cdot (\hat{\mathbf{u}} \times \partial_i \hat{\mathbf{u}}) \right] \right\} \, ,\nonumber\\
&\Rightarrow& \dot{\rho} = - \frac{1}{q} \partial_i J_i \, ,
\end{eqnarray}
where $J_i$ is the charge current density. This is nothing but the continuity equation.

Third, to obtain the equation of motion for $\hat{\mathbf{s}}$, by considering infinitesimal variations of the three vectors that maintain the orthonormality conditions,
\begin{eqnarray}
\hat{\mathbf{s}} &=& \hat{\mathbf{s}}_0 + \delta \hat{\mathbf{s}} = \hat{\mathbf{s}}_0 + a \hat{\mathbf{u}}_0 + b \hat{\mathbf{v}}_0 \, ,\nonumber\\
\hat{\mathbf{u}} &=& \hat{\mathbf{u}}_0 - a \hat{\mathbf{s}}_0 \, ,\nonumber\\
\hat{\mathbf{v}} &=& \hat{\mathbf{v}}_0 - b \hat{\mathbf{s}}_0 \, ,
\end{eqnarray}
we observe that, to zeroth order in $a$ and $b$,
\begin{eqnarray}
\frac{\delta}{\delta \hat{\mathbf{s}}} \left( \hat{\mathbf{s}} \cdot (\hat{\mathbf{u}} \times \partial_t \hat{\mathbf{u}}) \right) &=& \hat{\mathbf{u}}_0 \frac{\partial}{\partial a} \left( \hat{\mathbf{s}} \cdot (\hat{\mathbf{u}} \times \partial_t \hat{\mathbf{u}}) \right) + \hat{\mathbf{v}}_0 \frac{\partial}{\partial b} \left( \hat{\mathbf{s}} \cdot (\hat{\mathbf{u}} \times \partial_t \hat{\mathbf{u}}) \right) \, \nonumber\\
&=& - \hat{\mathbf{u}}_0 (\hat{\mathbf{s}}_0 \cdot (\hat{\mathbf{u}}_0 \times \partial_t \hat{\mathbf{s}}_0)) + \hat{\mathbf{v}}_0 (\hat{\mathbf{v}}_0 \cdot (\hat{\mathbf{u}}_0 \times \partial_t \hat{\mathbf{u}}_0)) \, \nonumber\\
&=& \hat{\mathbf{s}}_0 \times \partial_t \hat{\mathbf{s}}_0 \, .
\end{eqnarray}
Also, by the analogous steps,
\begin{equation}
\frac{\delta}{\delta \hat{\mathbf{s}}} \left( \hat{\mathbf{s}} \cdot (\hat{\mathbf{u}} \times \partial_i \hat{\mathbf{u}}) \right) = \hat{\mathbf{s}} \times \partial_i \hat{\mathbf{s}} \, .
\end{equation}

Then, from the Lagrangian, we obtain
\begin{eqnarray}
&& \frac{\delta L}{\delta \hat{\mathbf{s}}} = 0 \, , \nonumber\\
&\Rightarrow& \hbar \rho \hat{\mathbf{s}} \times \partial_t \hat{\mathbf{s}} = \frac{\delta F}{\delta \hat{\mathbf{s}}} \, , \nonumber\\
&\Rightarrow& \hbar \rho \partial_t \hat{\mathbf{s}} = - \frac{\hbar J_i}{q} \partial_i \hat{\mathbf{s}} + \partial_i [\rho A_s (\hat{\mathbf{s}} \times \partial_i \hat{\mathbf{s}})] + \rho D (\hat{\mathbf{s}} \cdot \hat{\mathbf{z}}) \hat{\mathbf{s}} \times \hat{\mathbf{z}} + \rho H \hat{\mathbf{s}} \times \hat{\mathbf{z}} \, . 
\end{eqnarray}

By using $\dot{\rho} = - (\partial_i J_i) / q$, the last equation can be recast into
\begin{equation}
\partial_t (\hbar \rho \hat{\mathbf{s}}) = - \partial_i \left[ \frac{\hbar J_i}{q} \hat{\mathbf{s}} - [\rho A_s (\hat{\mathbf{s}} \times \partial_i \hat{\mathbf{s}})] \right] + \rho D (\hat{\mathbf{s}} \cdot \hat{\mathbf{z}}) \hat{\mathbf{s}} \times \hat{\mathbf{z}} + \rho H \hat{\mathbf{s}} \times \hat{\mathbf{z}} \, .
\end{equation}
The left-hand side is the spin density, $\mathbf{s} = \hbar \rho \hat{\mathbf{s}}$. By identifying the first term on the right-hand side as the negative divergence of the spin-current density, we obtain the expression for the spin current density:
\begin{equation}
\mathbf{J}_i^s = \frac{\hbar J_i}{q} \hat{\mathbf{s}} - \rho A_s (\hat{\mathbf{s}} \times \partial_i \hat{\mathbf{s}}) \, .
\end{equation}
The second and the third terms are from the spin anisotropy and an external field in the $z$ direction, which break the full SU(2) spin-rotational symmetry to the U(1) rotational symmetry about the $z$ axis.

\end{widetext}
\end{appendix}

\bibliography{stsc}

\begin{thebibliography}{55}%
\makeatletter
\providecommand \@ifxundefined [1]{%
 \@ifx{#1\undefined}
}%
\providecommand \@ifnum [1]{%
 \ifnum #1\expandafter \@firstoftwo
 \else \expandafter \@secondoftwo
 \fi
}%
\providecommand \@ifx [1]{%
 \ifx #1\expandafter \@firstoftwo
 \else \expandafter \@secondoftwo
 \fi
}%
\providecommand \natexlab [1]{#1}%
\providecommand \enquote  [1]{``#1''}%
\providecommand \bibnamefont  [1]{#1}%
\providecommand \bibfnamefont [1]{#1}%
\providecommand \citenamefont [1]{#1}%
\providecommand \href@noop [0]{\@secondoftwo}%
\providecommand \href [0]{\begingroup \@sanitize@url \@href}%
\providecommand \@href[1]{\@@startlink{#1}\@@href}%
\providecommand \@@href[1]{\endgroup#1\@@endlink}%
\providecommand \@sanitize@url [0]{\catcode `\\12\catcode `\$12\catcode
  `\&12\catcode `\#12\catcode `\^12\catcode `\_12\catcode `\%12\relax}%
\providecommand \@@startlink[1]{}%
\providecommand \@@endlink[0]{}%
\providecommand \url  [0]{\begingroup\@sanitize@url \@url }%
\providecommand \@url [1]{\endgroup\@href {#1}{\urlprefix }}%
\providecommand \urlprefix  [0]{URL }%
\providecommand \Eprint [0]{\href }%
\providecommand \doibase [0]{https://doi.org/}%
\providecommand \selectlanguage [0]{\@gobble}%
\providecommand \bibinfo  [0]{\@secondoftwo}%
\providecommand \bibfield  [0]{\@secondoftwo}%
\providecommand \translation [1]{[#1]}%
\providecommand \BibitemOpen [0]{}%
\providecommand \bibitemStop [0]{}%
\providecommand \bibitemNoStop [0]{.\EOS\space}%
\providecommand \EOS [0]{\spacefactor3000\relax}%
\providecommand \BibitemShut  [1]{\csname bibitem#1\endcsname}%
\let\auto@bib@innerbib\@empty
\bibitem [{\citenamefont {Saxena}\ \emph {et~al.}(2000)\citenamefont {Saxena},
  \citenamefont {Agarwal}, \citenamefont {Ahilan}, \citenamefont {Grosche},
  \citenamefont {Haselwimmer}, \citenamefont {Steiner}, \citenamefont {Pugh},
  \citenamefont {Walker}, \citenamefont {Julian}, \citenamefont {Monthoux},
  \citenamefont {Lonzarich}, \citenamefont {Huxley}, \citenamefont {Sheikin},
  \citenamefont {Braithwaite},\ and\ \citenamefont
  {Flouquet}}]{SaxenaNature2000}%
  \BibitemOpen
  \bibfield  {author} {\bibinfo {author} {\bibfnamefont {S.~S.}\ \bibnamefont
  {Saxena}}, \bibinfo {author} {\bibfnamefont {P.}~\bibnamefont {Agarwal}},
  \bibinfo {author} {\bibfnamefont {K.}~\bibnamefont {Ahilan}}, \bibinfo
  {author} {\bibfnamefont {F.~M.}\ \bibnamefont {Grosche}}, \bibinfo {author}
  {\bibfnamefont {R.~K.~W.}\ \bibnamefont {Haselwimmer}}, \bibinfo {author}
  {\bibfnamefont {M.~J.}\ \bibnamefont {Steiner}}, \bibinfo {author}
  {\bibfnamefont {E.}~\bibnamefont {Pugh}}, \bibinfo {author} {\bibfnamefont
  {I.~R.}\ \bibnamefont {Walker}}, \bibinfo {author} {\bibfnamefont {S.~R.}\
  \bibnamefont {Julian}}, \bibinfo {author} {\bibfnamefont {P.}~\bibnamefont
  {Monthoux}}, \bibinfo {author} {\bibfnamefont {G.~G.}\ \bibnamefont
  {Lonzarich}}, \bibinfo {author} {\bibfnamefont {A.}~\bibnamefont {Huxley}},
  \bibinfo {author} {\bibfnamefont {I.}~\bibnamefont {Sheikin}}, \bibinfo
  {author} {\bibfnamefont {D.}~\bibnamefont {Braithwaite}},\ and\ \bibinfo
  {author} {\bibfnamefont {J.}~\bibnamefont {Flouquet}},\ }\bibfield  {title}
  {\bibinfo {title} {Superconductivity on the border of itinerant-electron
  ferromagnetism in uge2},\ }\href {https://doi.org/10.1038/35020500}
  {\bibfield  {journal} {\bibinfo  {journal} {Nature}\ }\textbf {\bibinfo
  {volume} {406}},\ \bibinfo {pages} {587} (\bibinfo {year}
  {2000})}\BibitemShut {NoStop}%
\bibitem [{\citenamefont {Aoki}\ \emph {et~al.}(2001)\citenamefont {Aoki},
  \citenamefont {Huxley}, \citenamefont {Ressouche}, \citenamefont
  {Braithwaite}, \citenamefont {Flouquet}, \citenamefont {Brison},
  \citenamefont {Lhotel},\ and\ \citenamefont {Paulsen}}]{AokiNature2001}%
  \BibitemOpen
  \bibfield  {author} {\bibinfo {author} {\bibfnamefont {D.}~\bibnamefont
  {Aoki}}, \bibinfo {author} {\bibfnamefont {A.}~\bibnamefont {Huxley}},
  \bibinfo {author} {\bibfnamefont {E.}~\bibnamefont {Ressouche}}, \bibinfo
  {author} {\bibfnamefont {D.}~\bibnamefont {Braithwaite}}, \bibinfo {author}
  {\bibfnamefont {J.}~\bibnamefont {Flouquet}}, \bibinfo {author}
  {\bibfnamefont {J.-P.}\ \bibnamefont {Brison}}, \bibinfo {author}
  {\bibfnamefont {E.}~\bibnamefont {Lhotel}},\ and\ \bibinfo {author}
  {\bibfnamefont {C.}~\bibnamefont {Paulsen}},\ }\bibfield  {title} {\bibinfo
  {title} {Coexistence of superconductivity and ferromagnetism in urhge},\
  }\href {https://doi.org/10.1038/35098048} {\bibfield  {journal} {\bibinfo
  {journal} {Nature}\ }\textbf {\bibinfo {volume} {413}},\ \bibinfo {pages}
  {613} (\bibinfo {year} {2001})}\BibitemShut {NoStop}%
\bibitem [{\citenamefont {Aoki}\ \emph {et~al.}(2019)\citenamefont {Aoki},
  \citenamefont {Ishida},\ and\ \citenamefont {Flouquet}}]{AokiJPSJ2019}%
  \BibitemOpen
  \bibfield  {author} {\bibinfo {author} {\bibfnamefont {D.}~\bibnamefont
  {Aoki}}, \bibinfo {author} {\bibfnamefont {K.}~\bibnamefont {Ishida}},\ and\
  \bibinfo {author} {\bibfnamefont {J.}~\bibnamefont {Flouquet}},\ }\bibfield
  {title} {\bibinfo {title} {{Review of U-based Ferromagnetic Superconductors:
  Comparison between UGe2, URhGe, and UCoGe}},\ }\href
  {https://doi.org/10.7566/JPSJ.88.022001} {\bibfield  {journal} {\bibinfo
  {journal} {J. Phys. Soc. Jpn.}\ }\textbf {\bibinfo {volume} {88}},\ \bibinfo
  {pages} {022001} (\bibinfo {year} {2019})}\BibitemShut {NoStop}%
\bibitem [{\citenamefont {Chen}\ \emph {et~al.}(2019)\citenamefont {Chen},
  \citenamefont {Jiang}, \citenamefont {Wu}, \citenamefont {Lyu}, \citenamefont
  {Li}, \citenamefont {Chittari}, \citenamefont {Watanabe}, \citenamefont
  {Taniguchi}, \citenamefont {Shi}, \citenamefont {Jung}, \citenamefont
  {Zhang},\ and\ \citenamefont {Wang}}]{ChenNP2019}%
  \BibitemOpen
  \bibfield  {author} {\bibinfo {author} {\bibfnamefont {G.}~\bibnamefont
  {Chen}}, \bibinfo {author} {\bibfnamefont {L.}~\bibnamefont {Jiang}},
  \bibinfo {author} {\bibfnamefont {S.}~\bibnamefont {Wu}}, \bibinfo {author}
  {\bibfnamefont {B.}~\bibnamefont {Lyu}}, \bibinfo {author} {\bibfnamefont
  {H.}~\bibnamefont {Li}}, \bibinfo {author} {\bibfnamefont {B.~L.}\
  \bibnamefont {Chittari}}, \bibinfo {author} {\bibfnamefont {K.}~\bibnamefont
  {Watanabe}}, \bibinfo {author} {\bibfnamefont {T.}~\bibnamefont {Taniguchi}},
  \bibinfo {author} {\bibfnamefont {Z.}~\bibnamefont {Shi}}, \bibinfo {author}
  {\bibfnamefont {J.}~\bibnamefont {Jung}}, \bibinfo {author} {\bibfnamefont
  {Y.}~\bibnamefont {Zhang}},\ and\ \bibinfo {author} {\bibfnamefont
  {F.}~\bibnamefont {Wang}},\ }\bibfield  {title} {\bibinfo {title} {Evidence
  of a gate-tunable mott insulator in a trilayer graphene
  moir{\'e}superlattice},\ }\href {https://doi.org/10.1038/s41567-018-0387-2}
  {\bibfield  {journal} {\bibinfo  {journal} {Nat. Phys.}\ }\textbf {\bibinfo
  {volume} {15}},\ \bibinfo {pages} {237} (\bibinfo {year} {2019})}\BibitemShut
  {NoStop}%
\bibitem [{\citenamefont {Sharpe}\ \emph {et~al.}(2019)\citenamefont {Sharpe},
  \citenamefont {Fox}, \citenamefont {Barnard}, \citenamefont {Finney},
  \citenamefont {Watanabe}, \citenamefont {Taniguchi}, \citenamefont
  {Kastner},\ and\ \citenamefont {Goldhaber-Gordon}}]{SharpeScience2019}%
  \BibitemOpen
  \bibfield  {author} {\bibinfo {author} {\bibfnamefont {A.~L.}\ \bibnamefont
  {Sharpe}}, \bibinfo {author} {\bibfnamefont {E.~J.}\ \bibnamefont {Fox}},
  \bibinfo {author} {\bibfnamefont {A.~W.}\ \bibnamefont {Barnard}}, \bibinfo
  {author} {\bibfnamefont {J.}~\bibnamefont {Finney}}, \bibinfo {author}
  {\bibfnamefont {K.}~\bibnamefont {Watanabe}}, \bibinfo {author}
  {\bibfnamefont {T.}~\bibnamefont {Taniguchi}}, \bibinfo {author}
  {\bibfnamefont {M.~A.}\ \bibnamefont {Kastner}},\ and\ \bibinfo {author}
  {\bibfnamefont {D.}~\bibnamefont {Goldhaber-Gordon}},\ }\bibfield  {title}
  {\bibinfo {title} {Emergent ferromagnetism near three-quarters filling in
  twisted bilayer graphene},\ }\href {https://doi.org/10.1126/science.aaw3780}
  {\bibfield  {journal} {\bibinfo  {journal} {Science}\ }\textbf {\bibinfo
  {volume} {365}},\ \bibinfo {pages} {605} (\bibinfo {year}
  {2019})}\BibitemShut {NoStop}%
\bibitem [{\citenamefont {Zondiner}\ \emph {et~al.}(2020)\citenamefont
  {Zondiner}, \citenamefont {Rozen}, \citenamefont {Rodan-Legrain},
  \citenamefont {Cao}, \citenamefont {Queiroz}, \citenamefont {Taniguchi},
  \citenamefont {Watanabe}, \citenamefont {Oreg}, \citenamefont {von Oppen},
  \citenamefont {Stern}, \citenamefont {Berg}, \citenamefont
  {Jarillo-Herrero},\ and\ \citenamefont {Ilani}}]{ZondinerNature2020}%
  \BibitemOpen
  \bibfield  {author} {\bibinfo {author} {\bibfnamefont {U.}~\bibnamefont
  {Zondiner}}, \bibinfo {author} {\bibfnamefont {A.}~\bibnamefont {Rozen}},
  \bibinfo {author} {\bibfnamefont {D.}~\bibnamefont {Rodan-Legrain}}, \bibinfo
  {author} {\bibfnamefont {Y.}~\bibnamefont {Cao}}, \bibinfo {author}
  {\bibfnamefont {R.}~\bibnamefont {Queiroz}}, \bibinfo {author} {\bibfnamefont
  {T.}~\bibnamefont {Taniguchi}}, \bibinfo {author} {\bibfnamefont
  {K.}~\bibnamefont {Watanabe}}, \bibinfo {author} {\bibfnamefont
  {Y.}~\bibnamefont {Oreg}}, \bibinfo {author} {\bibfnamefont {F.}~\bibnamefont
  {von Oppen}}, \bibinfo {author} {\bibfnamefont {A.}~\bibnamefont {Stern}},
  \bibinfo {author} {\bibfnamefont {E.}~\bibnamefont {Berg}}, \bibinfo {author}
  {\bibfnamefont {P.}~\bibnamefont {Jarillo-Herrero}},\ and\ \bibinfo {author}
  {\bibfnamefont {S.}~\bibnamefont {Ilani}},\ }\bibfield  {title} {\bibinfo
  {title} {Cascade of phase transitions and dirac revivals in magic-angle
  graphene},\ }\href {https://doi.org/10.1038/s41586-020-2373-y} {\bibfield
  {journal} {\bibinfo  {journal} {Nature}\ }\textbf {\bibinfo {volume} {582}},\
  \bibinfo {pages} {203} (\bibinfo {year} {2020})}\BibitemShut {NoStop}%
\bibitem [{\citenamefont {Vollhardt}\ and\ \citenamefont
  {Wolfle}(2013)}]{Vollhardt2013helium}%
  \BibitemOpen
  \bibfield  {author} {\bibinfo {author} {\bibfnamefont {D.}~\bibnamefont
  {Vollhardt}}\ and\ \bibinfo {author} {\bibfnamefont {P.}~\bibnamefont
  {Wolfle}},\ }\href@noop {} {\emph {\bibinfo {title} {The Superfluid Phases of
  Helium 3}}},\ Dover Books on Physics\ (\bibinfo  {publisher} {Dover
  Publications},\ \bibinfo {year} {2013})\BibitemShut {NoStop}%
\bibitem [{\citenamefont {Gilbert}(2004)}]{GilbertIEEE2004}%
  \BibitemOpen
  \bibfield  {author} {\bibinfo {author} {\bibfnamefont {T.}~\bibnamefont
  {Gilbert}},\ }\bibfield  {title} {\bibinfo {title} {A phenomenological theory
  of damping in ferromagnetic materials},\ }\href
  {https://doi.org/10.1109/TMAG.2004.836740} {\bibfield  {journal} {\bibinfo
  {journal} {IEEE Trans. Magn.}\ }\textbf {\bibinfo {volume} {40}},\ \bibinfo
  {pages} {3443} (\bibinfo {year} {2004})}\BibitemShut {NoStop}%
\bibitem [{\citenamefont {Tserkovnyak}\ \emph {et~al.}(2008)\citenamefont
  {Tserkovnyak}, \citenamefont {Brataas},\ and\ \citenamefont
  {Bauer}}]{TserkovnyakJMMM2008}%
  \BibitemOpen
  \bibfield  {author} {\bibinfo {author} {\bibfnamefont {Y.}~\bibnamefont
  {Tserkovnyak}}, \bibinfo {author} {\bibfnamefont {A.}~\bibnamefont
  {Brataas}},\ and\ \bibinfo {author} {\bibfnamefont {G.~E.}\ \bibnamefont
  {Bauer}},\ }\bibfield  {title} {\bibinfo {title} {Theory of current-driven
  magnetization dynamics in inhomogeneous ferromagnets},\ }\href
  {https://doi.org/http://dx.doi.org/10.1016/j.jmmm.2007.12.012} {\bibfield
  {journal} {\bibinfo  {journal} {J. Magn. Magn. Mater.}\ }\textbf {\bibinfo
  {volume} {320}},\ \bibinfo {pages} {1282 } (\bibinfo {year}
  {2008})}\BibitemShut {NoStop}%
\bibitem [{\citenamefont {Halperin}\ and\ \citenamefont
  {Nelson}(1979)}]{HalperinJLTP1979}%
  \BibitemOpen
  \bibfield  {author} {\bibinfo {author} {\bibfnamefont {B.~I.}\ \bibnamefont
  {Halperin}}\ and\ \bibinfo {author} {\bibfnamefont {D.~R.}\ \bibnamefont
  {Nelson}},\ }\bibfield  {title} {\bibinfo {title} {Resistive transition in
  superconducting films},\ }\href {https://doi.org/10.1007/BF00116988}
  {\bibfield  {journal} {\bibinfo  {journal} {J. Low Temp. Phys.}\ }\textbf
  {\bibinfo {volume} {36}},\ \bibinfo {pages} {599} (\bibinfo {year}
  {1979})}\BibitemShut {NoStop}%
\bibitem [{\citenamefont {Tinkham}(2004)}]{Tinkham2004}%
  \BibitemOpen
  \bibfield  {author} {\bibinfo {author} {\bibfnamefont {M.}~\bibnamefont
  {Tinkham}},\ }\href@noop {} {\emph {\bibinfo {title} {Introduction to
  Superconductivity}}}\ (\bibinfo  {publisher} {Dover, New York},\ \bibinfo
  {year} {2004})\BibitemShut {NoStop}%
\bibitem [{\citenamefont {Cornfeld}\ \emph {et~al.}(2021)\citenamefont
  {Cornfeld}, \citenamefont {Rudner},\ and\ \citenamefont
  {Berg}}]{CornfeldPRR2021}%
  \BibitemOpen
  \bibfield  {author} {\bibinfo {author} {\bibfnamefont {E.}~\bibnamefont
  {Cornfeld}}, \bibinfo {author} {\bibfnamefont {M.~S.}\ \bibnamefont
  {Rudner}},\ and\ \bibinfo {author} {\bibfnamefont {E.}~\bibnamefont {Berg}},\
  }\bibfield  {title} {\bibinfo {title} {Spin-polarized superconductivity:
  Order parameter topology, current dissipation, and multiple-period josephson
  effect},\ }\href {https://doi.org/10.1103/PhysRevResearch.3.013051}
  {\bibfield  {journal} {\bibinfo  {journal} {Phys. Rev. Research}\ }\textbf
  {\bibinfo {volume} {3}},\ \bibinfo {pages} {013051} (\bibinfo {year}
  {2021})}\BibitemShut {NoStop}%
\bibitem [{\citenamefont {Poniatowski}\ \emph {et~al.}(2022)\citenamefont
  {Poniatowski}, \citenamefont {Curtis}, \citenamefont {B\o{}ttcher},
  \citenamefont {Galitski}, \citenamefont {Yacoby}, \citenamefont {Narang},\
  and\ \citenamefont {Demler}}]{PoniatowskiPRL2022}%
  \BibitemOpen
  \bibfield  {author} {\bibinfo {author} {\bibfnamefont {N.~R.}\ \bibnamefont
  {Poniatowski}}, \bibinfo {author} {\bibfnamefont {J.~B.}\ \bibnamefont
  {Curtis}}, \bibinfo {author} {\bibfnamefont {C.~G.~L.}\ \bibnamefont
  {B\o{}ttcher}}, \bibinfo {author} {\bibfnamefont {V.~M.}\ \bibnamefont
  {Galitski}}, \bibinfo {author} {\bibfnamefont {A.}~\bibnamefont {Yacoby}},
  \bibinfo {author} {\bibfnamefont {P.}~\bibnamefont {Narang}},\ and\ \bibinfo
  {author} {\bibfnamefont {E.}~\bibnamefont {Demler}},\ }\bibfield  {title}
  {\bibinfo {title} {Surface cooper-pair spin waves in triplet
  superconductors},\ }\href {https://doi.org/10.1103/PhysRevLett.129.237002}
  {\bibfield  {journal} {\bibinfo  {journal} {Phys. Rev. Lett.}\ }\textbf
  {\bibinfo {volume} {129}},\ \bibinfo {pages} {237002} (\bibinfo {year}
  {2022})}\BibitemShut {NoStop}%
\bibitem [{\citenamefont {Volovik}(1987)}]{VolovikJPC1987}%
  \BibitemOpen
  \bibfield  {author} {\bibinfo {author} {\bibfnamefont {G.~E.}\ \bibnamefont
  {Volovik}},\ }\bibfield  {title} {\bibinfo {title} {Linear momentum in
  ferromagnets},\ }\href {http://stacks.iop.org/0022-3719/20/i=7/a=003}
  {\bibfield  {journal} {\bibinfo  {journal} {J. Phys. C: Solid State Phys.}\
  }\textbf {\bibinfo {volume} {20}},\ \bibinfo {pages} {L83} (\bibinfo {year}
  {1987})}\BibitemShut {NoStop}%
\bibitem [{\citenamefont {Wong}\ and\ \citenamefont
  {Tserkovnyak}(2009)}]{WongPRB2009}%
  \BibitemOpen
  \bibfield  {author} {\bibinfo {author} {\bibfnamefont {C.~H.}\ \bibnamefont
  {Wong}}\ and\ \bibinfo {author} {\bibfnamefont {Y.}~\bibnamefont
  {Tserkovnyak}},\ }\bibfield  {title} {\bibinfo {title} {Hydrodynamic theory
  of coupled current and magnetization dynamics in spin-textured
  ferromagnets},\ }\href {https://doi.org/10.1103/PhysRevB.80.184411}
  {\bibfield  {journal} {\bibinfo  {journal} {Phys. Rev. B}\ }\textbf {\bibinfo
  {volume} {80}},\ \bibinfo {pages} {184411} (\bibinfo {year}
  {2009})}\BibitemShut {NoStop}%
\bibitem [{\citenamefont {Zang}\ \emph {et~al.}(2011)\citenamefont {Zang},
  \citenamefont {Mostovoy}, \citenamefont {Han},\ and\ \citenamefont
  {Nagaosa}}]{ZangPRL2011}%
  \BibitemOpen
  \bibfield  {author} {\bibinfo {author} {\bibfnamefont {J.}~\bibnamefont
  {Zang}}, \bibinfo {author} {\bibfnamefont {M.}~\bibnamefont {Mostovoy}},
  \bibinfo {author} {\bibfnamefont {J.~H.}\ \bibnamefont {Han}},\ and\ \bibinfo
  {author} {\bibfnamefont {N.}~\bibnamefont {Nagaosa}},\ }\bibfield  {title}
  {\bibinfo {title} {Dynamics of skyrmion crystals in metallic thin films},\
  }\href {https://doi.org/10.1103/PhysRevLett.107.136804} {\bibfield  {journal}
  {\bibinfo  {journal} {Phys. Rev. Lett.}\ }\textbf {\bibinfo {volume} {107}},\
  \bibinfo {pages} {136804} (\bibinfo {year} {2011})}\BibitemShut {NoStop}%
\bibitem [{\citenamefont {Nagaosa}\ and\ \citenamefont
  {Tokura}(2013)}]{NagaosaNN2013}%
  \BibitemOpen
  \bibfield  {author} {\bibinfo {author} {\bibfnamefont {N.}~\bibnamefont
  {Nagaosa}}\ and\ \bibinfo {author} {\bibfnamefont {Y.}~\bibnamefont
  {Tokura}},\ }\bibfield  {title} {\bibinfo {title} {Topological properties and
  dynamics of magnetic skyrmions},\ }\href
  {http://dx.doi.org/10.1038/nnano.2013.243} {\bibfield  {journal} {\bibinfo
  {journal} {Nat. Nanotechnol.}\ }\textbf {\bibinfo {volume} {8}},\ \bibinfo
  {pages} {899} (\bibinfo {year} {2013})}\BibitemShut {NoStop}%
\bibitem [{\citenamefont {Yang}\ \emph {et~al.}(2009)\citenamefont {Yang},
  \citenamefont {Beach}, \citenamefont {Knutson}, \citenamefont {Xiao},
  \citenamefont {Niu}, \citenamefont {Tsoi},\ and\ \citenamefont
  {Erskine}}]{YangPRL2009}%
  \BibitemOpen
  \bibfield  {author} {\bibinfo {author} {\bibfnamefont {S.~A.}\ \bibnamefont
  {Yang}}, \bibinfo {author} {\bibfnamefont {G.~S.~D.}\ \bibnamefont {Beach}},
  \bibinfo {author} {\bibfnamefont {C.}~\bibnamefont {Knutson}}, \bibinfo
  {author} {\bibfnamefont {D.}~\bibnamefont {Xiao}}, \bibinfo {author}
  {\bibfnamefont {Q.}~\bibnamefont {Niu}}, \bibinfo {author} {\bibfnamefont
  {M.}~\bibnamefont {Tsoi}},\ and\ \bibinfo {author} {\bibfnamefont {J.~L.}\
  \bibnamefont {Erskine}},\ }\bibfield  {title} {\bibinfo {title} {Universal
  electromotive force induced by domain wall motion},\ }\href
  {https://doi.org/10.1103/PhysRevLett.102.067201} {\bibfield  {journal}
  {\bibinfo  {journal} {Phys. Rev. Lett.}\ }\textbf {\bibinfo {volume} {102}},\
  \bibinfo {pages} {067201} (\bibinfo {year} {2009})}\BibitemShut {NoStop}%
\bibitem [{\citenamefont {Neubauer}\ \emph {et~al.}(2009)\citenamefont
  {Neubauer}, \citenamefont {Pfleiderer}, \citenamefont {Binz}, \citenamefont
  {Rosch}, \citenamefont {Ritz}, \citenamefont {Niklowitz},\ and\ \citenamefont
  {B\"oni}}]{NeubauerPRL2009}%
  \BibitemOpen
  \bibfield  {author} {\bibinfo {author} {\bibfnamefont {A.}~\bibnamefont
  {Neubauer}}, \bibinfo {author} {\bibfnamefont {C.}~\bibnamefont
  {Pfleiderer}}, \bibinfo {author} {\bibfnamefont {B.}~\bibnamefont {Binz}},
  \bibinfo {author} {\bibfnamefont {A.}~\bibnamefont {Rosch}}, \bibinfo
  {author} {\bibfnamefont {R.}~\bibnamefont {Ritz}}, \bibinfo {author}
  {\bibfnamefont {P.~G.}\ \bibnamefont {Niklowitz}},\ and\ \bibinfo {author}
  {\bibfnamefont {P.}~\bibnamefont {B\"oni}},\ }\bibfield  {title} {\bibinfo
  {title} {{Topological Hall Effect in the $A$ Phase of MnSi}},\ }\href
  {https://doi.org/10.1103/PhysRevLett.102.186602} {\bibfield  {journal}
  {\bibinfo  {journal} {Phys. Rev. Lett.}\ }\textbf {\bibinfo {volume} {102}},\
  \bibinfo {pages} {186602} (\bibinfo {year} {2009})}\BibitemShut {NoStop}%
\bibitem [{\citenamefont {Yang}\ \emph {et~al.}(2010)\citenamefont {Yang},
  \citenamefont {Beach}, \citenamefont {Knutson}, \citenamefont {Xiao},
  \citenamefont {Zhang}, \citenamefont {Tsoi}, \citenamefont {Niu},
  \citenamefont {MacDonald},\ and\ \citenamefont {Erskine}}]{YangPRB2010}%
  \BibitemOpen
  \bibfield  {author} {\bibinfo {author} {\bibfnamefont {S.~A.}\ \bibnamefont
  {Yang}}, \bibinfo {author} {\bibfnamefont {G.~S.~D.}\ \bibnamefont {Beach}},
  \bibinfo {author} {\bibfnamefont {C.}~\bibnamefont {Knutson}}, \bibinfo
  {author} {\bibfnamefont {D.}~\bibnamefont {Xiao}}, \bibinfo {author}
  {\bibfnamefont {Z.}~\bibnamefont {Zhang}}, \bibinfo {author} {\bibfnamefont
  {M.}~\bibnamefont {Tsoi}}, \bibinfo {author} {\bibfnamefont {Q.}~\bibnamefont
  {Niu}}, \bibinfo {author} {\bibfnamefont {A.~H.}\ \bibnamefont {MacDonald}},\
  and\ \bibinfo {author} {\bibfnamefont {J.~L.}\ \bibnamefont {Erskine}},\
  }\bibfield  {title} {\bibinfo {title} {Topological electromotive force from
  domain-wall dynamics in a ferromagnet},\ }\href
  {https://doi.org/10.1103/PhysRevB.82.054410} {\bibfield  {journal} {\bibinfo
  {journal} {Phys. Rev. B}\ }\textbf {\bibinfo {volume} {82}},\ \bibinfo
  {pages} {054410} (\bibinfo {year} {2010})}\BibitemShut {NoStop}%
\bibitem [{\citenamefont {Bisig}\ \emph {et~al.}(2016)\citenamefont {Bisig},
  \citenamefont {Akosa}, \citenamefont {Moon}, \citenamefont {Rhensius},
  \citenamefont {Moutafis}, \citenamefont {von Bieren}, \citenamefont
  {Heidler}, \citenamefont {Kiliani}, \citenamefont {Kammerer}, \citenamefont
  {Curcic}, \citenamefont {Weigand}, \citenamefont {Tyliszczak}, \citenamefont
  {Van~Waeyenberge}, \citenamefont {Stoll}, \citenamefont {Sch\"utz},
  \citenamefont {Lee}, \citenamefont {Manchon},\ and\ \citenamefont
  {Kl\"aui}}]{BisigPRL2016}%
  \BibitemOpen
  \bibfield  {author} {\bibinfo {author} {\bibfnamefont {A.}~\bibnamefont
  {Bisig}}, \bibinfo {author} {\bibfnamefont {C.~A.}\ \bibnamefont {Akosa}},
  \bibinfo {author} {\bibfnamefont {J.-H.}\ \bibnamefont {Moon}}, \bibinfo
  {author} {\bibfnamefont {J.}~\bibnamefont {Rhensius}}, \bibinfo {author}
  {\bibfnamefont {C.}~\bibnamefont {Moutafis}}, \bibinfo {author}
  {\bibfnamefont {A.}~\bibnamefont {von Bieren}}, \bibinfo {author}
  {\bibfnamefont {J.}~\bibnamefont {Heidler}}, \bibinfo {author} {\bibfnamefont
  {G.}~\bibnamefont {Kiliani}}, \bibinfo {author} {\bibfnamefont
  {M.}~\bibnamefont {Kammerer}}, \bibinfo {author} {\bibfnamefont
  {M.}~\bibnamefont {Curcic}}, \bibinfo {author} {\bibfnamefont
  {M.}~\bibnamefont {Weigand}}, \bibinfo {author} {\bibfnamefont
  {T.}~\bibnamefont {Tyliszczak}}, \bibinfo {author} {\bibfnamefont
  {B.}~\bibnamefont {Van~Waeyenberge}}, \bibinfo {author} {\bibfnamefont
  {H.}~\bibnamefont {Stoll}}, \bibinfo {author} {\bibfnamefont
  {G.}~\bibnamefont {Sch\"utz}}, \bibinfo {author} {\bibfnamefont {K.-J.}\
  \bibnamefont {Lee}}, \bibinfo {author} {\bibfnamefont {A.}~\bibnamefont
  {Manchon}},\ and\ \bibinfo {author} {\bibfnamefont {M.}~\bibnamefont
  {Kl\"aui}},\ }\bibfield  {title} {\bibinfo {title} {{Enhanced Nonadiabaticity
  in Vortex Cores due to the Emergent Hall Effect}},\ }\href
  {https://doi.org/10.1103/PhysRevLett.117.277203} {\bibfield  {journal}
  {\bibinfo  {journal} {Phys. Rev. Lett.}\ }\textbf {\bibinfo {volume} {117}},\
  \bibinfo {pages} {277203} (\bibinfo {year} {2016})}\BibitemShut {NoStop}%
\bibitem [{\citenamefont {Cross}(1975)}]{Cross1975}%
  \BibitemOpen
  \bibfield  {author} {\bibinfo {author} {\bibfnamefont {M.~C.}\ \bibnamefont
  {Cross}},\ }\bibfield  {title} {\bibinfo {title} {A generalized
  ginzburg-landau approach to the superfluidity of helium 3},\ }\href
  {https://doi.org/10.1007/BF01141607} {\bibfield  {journal} {\bibinfo
  {journal} {J. Low Temp. Phys.}\ }\textbf {\bibinfo {volume} {21}},\ \bibinfo
  {pages} {525} (\bibinfo {year} {1975})}\BibitemShut {NoStop}%
\bibitem [{\citenamefont {Mermin}(1978)}]{Mermin1978}%
  \BibitemOpen
  \bibfield  {author} {\bibinfo {author} {\bibfnamefont {N.~D.}\ \bibnamefont
  {Mermin}},\ }\bibfield  {title} {\bibinfo {title} {Superfluidity in
  helium-3},\ }in\ \href
  {https://doi.org/https://doi.org/10.1016/B978-0-444-85117-8.50010-6} {\emph
  {\bibinfo {booktitle} {Quantum Liquids}}},\ \bibinfo {editor} {edited by\
  \bibinfo {editor} {\bibfnamefont {J.}~\bibnamefont {Ruvalds}}\ and\ \bibinfo
  {editor} {\bibfnamefont {T.}~\bibnamefont {Regge}}}\ (\bibinfo  {publisher}
  {Elsevier},\ \bibinfo {year} {1978})\ pp.\ \bibinfo {pages}
  {195--226}\BibitemShut {NoStop}%
\bibitem [{\citenamefont {Jiang}\ \emph {et~al.}(2020)\citenamefont {Jiang},
  \citenamefont {Hansson},\ and\ \citenamefont {Wilczek}}]{QDJiang2020}%
  \BibitemOpen
  \bibfield  {author} {\bibinfo {author} {\bibfnamefont {Q.-D.}\ \bibnamefont
  {Jiang}}, \bibinfo {author} {\bibfnamefont {T.~H.}\ \bibnamefont {Hansson}},\
  and\ \bibinfo {author} {\bibfnamefont {F.}~\bibnamefont {Wilczek}},\
  }\bibfield  {title} {\bibinfo {title} {Geometric induction in chiral
  superconductors},\ }\href {https://doi.org/10.1103/PhysRevLett.124.197001}
  {\bibfield  {journal} {\bibinfo  {journal} {Phys. Rev. Lett.}\ }\textbf
  {\bibinfo {volume} {124}},\ \bibinfo {pages} {197001} (\bibinfo {year}
  {2020})}\BibitemShut {NoStop}%
\bibitem [{\citenamefont {Jiang}\ and\ \citenamefont
  {Balatsky}(2022)}]{QDJiang2022}%
  \BibitemOpen
  \bibfield  {author} {\bibinfo {author} {\bibfnamefont {Q.-D.}\ \bibnamefont
  {Jiang}}\ and\ \bibinfo {author} {\bibfnamefont {A.}~\bibnamefont
  {Balatsky}},\ }\bibfield  {title} {\bibinfo {title} {Geometric induction in
  chiral superfluids},\ }\href {https://doi.org/10.1103/PhysRevLett.129.016801}
  {\bibfield  {journal} {\bibinfo  {journal} {Phys. Rev. Lett.}\ }\textbf
  {\bibinfo {volume} {129}},\ \bibinfo {pages} {016801} (\bibinfo {year}
  {2022})}\BibitemShut {NoStop}%
\bibitem [{\citenamefont {Mermin}\ and\ \citenamefont
  {Ho}(1976)}]{MerminHo1976}%
  \BibitemOpen
  \bibfield  {author} {\bibinfo {author} {\bibfnamefont {N.~D.}\ \bibnamefont
  {Mermin}}\ and\ \bibinfo {author} {\bibfnamefont {T.-L.}\ \bibnamefont
  {Ho}},\ }\bibfield  {title} {\bibinfo {title} {Circulation and angular
  momentum in the $a$ phase of superfluid helium-3},\ }\href
  {https://doi.org/10.1103/PhysRevLett.36.594} {\bibfield  {journal} {\bibinfo
  {journal} {Phys. Rev. Lett.}\ }\textbf {\bibinfo {volume} {36}},\ \bibinfo
  {pages} {594} (\bibinfo {year} {1976})}\BibitemShut {NoStop}%
\bibitem [{\citenamefont {Volovik}(2003)}]{Volovik2003}%
  \BibitemOpen
  \bibfield  {author} {\bibinfo {author} {\bibfnamefont {G.}~\bibnamefont
  {Volovik}},\ }\href {https://books.google.co.kr/books?id=cbngYQWAiDEC} {\emph
  {\bibinfo {title} {The Universe in a Helium Droplet}}},\ International Series
  of Monographs on Physics\ (\bibinfo  {publisher} {Clarendon Press},\ \bibinfo
  {year} {2003})\BibitemShut {NoStop}%
\bibitem [{\citenamefont {Li}\ \emph {et~al.}(2009)\citenamefont {Li},
  \citenamefont {Toner},\ and\ \citenamefont {Belitz}}]{LiPRB2009}%
  \BibitemOpen
  \bibfield  {author} {\bibinfo {author} {\bibfnamefont {Q.}~\bibnamefont
  {Li}}, \bibinfo {author} {\bibfnamefont {J.}~\bibnamefont {Toner}},\ and\
  \bibinfo {author} {\bibfnamefont {D.}~\bibnamefont {Belitz}},\ }\bibfield
  {title} {\bibinfo {title} {Skyrmion versus vortex flux lattices in $p$-wave
  superconductors},\ }\href {https://doi.org/10.1103/PhysRevB.79.014517}
  {\bibfield  {journal} {\bibinfo  {journal} {Phys. Rev. B}\ }\textbf {\bibinfo
  {volume} {79}},\ \bibinfo {pages} {014517} (\bibinfo {year}
  {2009})}\BibitemShut {NoStop}%
\bibitem [{\citenamefont {Landau}\ \emph {et~al.}(1980)\citenamefont {Landau},
  \citenamefont {Lifshitz},\ and\ \citenamefont {Pitaevskii}}]{LL5}%
  \BibitemOpen
  \bibfield  {author} {\bibinfo {author} {\bibfnamefont {L.~D.}\ \bibnamefont
  {Landau}}, \bibinfo {author} {\bibfnamefont {E.~M.}\ \bibnamefont
  {Lifshitz}},\ and\ \bibinfo {author} {\bibfnamefont {L.~P.}\ \bibnamefont
  {Pitaevskii}},\ }\href@noop {} {\emph {\bibinfo {title} {Statistical Physics,
  Part 1}}},\ \bibinfo {edition} {3rd}\ ed.\ (\bibinfo  {publisher} {Pergamon
  Press, New York},\ \bibinfo {year} {1980})\BibitemShut {NoStop}%
\bibitem [{\citenamefont {Slonczewski}(1996)}]{SlonczewskiJMMM1996}%
  \BibitemOpen
  \bibfield  {author} {\bibinfo {author} {\bibfnamefont {J.}~\bibnamefont
  {Slonczewski}},\ }\bibfield  {title} {\bibinfo {title} {Current-driven
  excitation of magnetic multilayers},\ }\href
  {https://doi.org/http://dx.doi.org/10.1016/0304-8853(96)00062-5} {\bibfield
  {journal} {\bibinfo  {journal} {J. Magn. Magn. Mater.}\ }\textbf {\bibinfo
  {volume} {159}},\ \bibinfo {pages} {L1 } (\bibinfo {year}
  {1996})}\BibitemShut {NoStop}%
\bibitem [{\citenamefont {Berger}(1996)}]{BergerPRB1996}%
  \BibitemOpen
  \bibfield  {author} {\bibinfo {author} {\bibfnamefont {L.}~\bibnamefont
  {Berger}},\ }\bibfield  {title} {\bibinfo {title} {Emission of spin waves by
  a magnetic multilayer traversed by a current},\ }\href
  {https://doi.org/10.1103/PhysRevB.54.9353} {\bibfield  {journal} {\bibinfo
  {journal} {Phys. Rev. B}\ }\textbf {\bibinfo {volume} {54}},\ \bibinfo
  {pages} {9353} (\bibinfo {year} {1996})}\BibitemShut {NoStop}%
\bibitem [{\citenamefont {Barnett}\ \emph {et~al.}(2009)\citenamefont
  {Barnett}, \citenamefont {Podolsky},\ and\ \citenamefont
  {Refael}}]{BarnettPRB2009}%
  \BibitemOpen
  \bibfield  {author} {\bibinfo {author} {\bibfnamefont {R.}~\bibnamefont
  {Barnett}}, \bibinfo {author} {\bibfnamefont {D.}~\bibnamefont {Podolsky}},\
  and\ \bibinfo {author} {\bibfnamefont {G.}~\bibnamefont {Refael}},\
  }\bibfield  {title} {\bibinfo {title} {Geometrical approach to hydrodynamics
  and low-energy excitations of spinor condensates},\ }\href
  {https://doi.org/10.1103/PhysRevB.80.024420} {\bibfield  {journal} {\bibinfo
  {journal} {Phys. Rev. B}\ }\textbf {\bibinfo {volume} {80}},\ \bibinfo
  {pages} {024420} (\bibinfo {year} {2009})}\BibitemShut {NoStop}%
\bibitem [{\citenamefont {Tserkovnyak}\ \emph {et~al.}(2002)\citenamefont
  {Tserkovnyak}, \citenamefont {Brataas},\ and\ \citenamefont
  {Bauer}}]{TserkovnyakPRL2002}%
  \BibitemOpen
  \bibfield  {author} {\bibinfo {author} {\bibfnamefont {Y.}~\bibnamefont
  {Tserkovnyak}}, \bibinfo {author} {\bibfnamefont {A.}~\bibnamefont
  {Brataas}},\ and\ \bibinfo {author} {\bibfnamefont {G.~E.~W.}\ \bibnamefont
  {Bauer}},\ }\bibfield  {title} {\bibinfo {title} {{Enhanced Gilbert Damping
  in Thin Ferromagnetic Films}},\ }\href
  {https://doi.org/10.1103/PhysRevLett.88.117601} {\bibfield  {journal}
  {\bibinfo  {journal} {Phys. Rev. Lett.}\ }\textbf {\bibinfo {volume} {88}},\
  \bibinfo {pages} {117601} (\bibinfo {year} {2002})}\BibitemShut {NoStop}%
\bibitem [{\citenamefont {Hickey}\ and\ \citenamefont
  {Moodera}(2009)}]{HickeyPRL2009}%
  \BibitemOpen
  \bibfield  {author} {\bibinfo {author} {\bibfnamefont {M.~C.}\ \bibnamefont
  {Hickey}}\ and\ \bibinfo {author} {\bibfnamefont {J.~S.}\ \bibnamefont
  {Moodera}},\ }\bibfield  {title} {\bibinfo {title} {Origin of intrinsic
  gilbert damping},\ }\href {https://doi.org/10.1103/PhysRevLett.102.137601}
  {\bibfield  {journal} {\bibinfo  {journal} {Phys. Rev. Lett.}\ }\textbf
  {\bibinfo {volume} {102}},\ \bibinfo {pages} {137601} (\bibinfo {year}
  {2009})}\BibitemShut {NoStop}%
\bibitem [{\citenamefont {Starikov}\ \emph {et~al.}(2010)\citenamefont
  {Starikov}, \citenamefont {Kelly}, \citenamefont {Brataas}, \citenamefont
  {Tserkovnyak},\ and\ \citenamefont {Bauer}}]{StarikovPRL2010}%
  \BibitemOpen
  \bibfield  {author} {\bibinfo {author} {\bibfnamefont {A.~A.}\ \bibnamefont
  {Starikov}}, \bibinfo {author} {\bibfnamefont {P.~J.}\ \bibnamefont {Kelly}},
  \bibinfo {author} {\bibfnamefont {A.}~\bibnamefont {Brataas}}, \bibinfo
  {author} {\bibfnamefont {Y.}~\bibnamefont {Tserkovnyak}},\ and\ \bibinfo
  {author} {\bibfnamefont {G.~E.~W.}\ \bibnamefont {Bauer}},\ }\bibfield
  {title} {\bibinfo {title} {Unified first-principles study of gilbert damping,
  spin-flip diffusion, and resistivity in transition metal alloys},\ }\href
  {https://doi.org/10.1103/PhysRevLett.105.236601} {\bibfield  {journal}
  {\bibinfo  {journal} {Phys. Rev. Lett.}\ }\textbf {\bibinfo {volume} {105}},\
  \bibinfo {pages} {236601} (\bibinfo {year} {2010})}\BibitemShut {NoStop}%
\bibitem [{\citenamefont {R\"uckriegel}\ \emph {et~al.}(2014)\citenamefont
  {R\"uckriegel}, \citenamefont {Kopietz}, \citenamefont {Bozhko},
  \citenamefont {Serga},\ and\ \citenamefont
  {Hillebrands}}]{RuckriegelPRB2014}%
  \BibitemOpen
  \bibfield  {author} {\bibinfo {author} {\bibfnamefont {A.}~\bibnamefont
  {R\"uckriegel}}, \bibinfo {author} {\bibfnamefont {P.}~\bibnamefont
  {Kopietz}}, \bibinfo {author} {\bibfnamefont {D.~A.}\ \bibnamefont {Bozhko}},
  \bibinfo {author} {\bibfnamefont {A.~A.}\ \bibnamefont {Serga}},\ and\
  \bibinfo {author} {\bibfnamefont {B.}~\bibnamefont {Hillebrands}},\
  }\bibfield  {title} {\bibinfo {title} {Magnetoelastic modes and lifetime of
  magnons in thin yttrium iron garnet films},\ }\href
  {https://doi.org/10.1103/PhysRevB.89.184413} {\bibfield  {journal} {\bibinfo
  {journal} {Phys. Rev. B}\ }\textbf {\bibinfo {volume} {89}},\ \bibinfo
  {pages} {184413} (\bibinfo {year} {2014})}\BibitemShut {NoStop}%
\bibitem [{\citenamefont {Kim}\ \emph {et~al.}(2014)\citenamefont {Kim},
  \citenamefont {Tserkovnyak},\ and\ \citenamefont
  {Tchernyshyov}}]{KimPRB2014}%
  \BibitemOpen
  \bibfield  {author} {\bibinfo {author} {\bibfnamefont {S.~K.}\ \bibnamefont
  {Kim}}, \bibinfo {author} {\bibfnamefont {Y.}~\bibnamefont {Tserkovnyak}},\
  and\ \bibinfo {author} {\bibfnamefont {O.}~\bibnamefont {Tchernyshyov}},\
  }\bibfield  {title} {\bibinfo {title} {Propulsion of a domain wall in an
  antiferromagnet by magnons},\ }\href
  {https://doi.org/10.1103/PhysRevB.90.104406} {\bibfield  {journal} {\bibinfo
  {journal} {Phys. Rev. B}\ }\textbf {\bibinfo {volume} {90}},\ \bibinfo
  {pages} {104406} (\bibinfo {year} {2014})}\BibitemShut {NoStop}%
\bibitem [{\citenamefont {Dasgupta}\ and\ \citenamefont
  {Tchernyshyov}(2018)}]{DasguptaPRB2018}%
  \BibitemOpen
  \bibfield  {author} {\bibinfo {author} {\bibfnamefont {S.}~\bibnamefont
  {Dasgupta}}\ and\ \bibinfo {author} {\bibfnamefont {O.}~\bibnamefont
  {Tchernyshyov}},\ }\bibfield  {title} {\bibinfo {title} {Energy-momentum
  tensor of a ferromagnet},\ }\href
  {https://doi.org/10.1103/PhysRevB.98.224401} {\bibfield  {journal} {\bibinfo
  {journal} {Phys. Rev. B}\ }\textbf {\bibinfo {volume} {98}},\ \bibinfo
  {pages} {224401} (\bibinfo {year} {2018})}\BibitemShut {NoStop}%
\bibitem [{\citenamefont {Kamien}(2002)}]{KamienRMP2002}%
  \BibitemOpen
  \bibfield  {author} {\bibinfo {author} {\bibfnamefont {R.~D.}\ \bibnamefont
  {Kamien}},\ }\bibfield  {title} {\bibinfo {title} {{The geometry of soft
  materials: a primer}},\ }\href {https://doi.org/10.1103/RevModPhys.74.953}
  {\bibfield  {journal} {\bibinfo  {journal} {Rev. Mod. Phys.}\ }\textbf
  {\bibinfo {volume} {74}},\ \bibinfo {pages} {953} (\bibinfo {year}
  {2002})}\BibitemShut {NoStop}%
\bibitem [{\citenamefont {Takashima}\ \emph {et~al.}(2017)\citenamefont
  {Takashima}, \citenamefont {Fujimoto},\ and\ \citenamefont
  {Yokoyama}}]{Takashima2017}%
  \BibitemOpen
  \bibfield  {author} {\bibinfo {author} {\bibfnamefont {R.}~\bibnamefont
  {Takashima}}, \bibinfo {author} {\bibfnamefont {S.}~\bibnamefont
  {Fujimoto}},\ and\ \bibinfo {author} {\bibfnamefont {T.}~\bibnamefont
  {Yokoyama}},\ }\bibfield  {title} {\bibinfo {title} {Adiabatic and
  nonadiabatic spin torques induced by a spin-triplet supercurrent},\ }\href
  {https://doi.org/10.1103/PhysRevB.96.121203} {\bibfield  {journal} {\bibinfo
  {journal} {Phys. Rev. B}\ }\textbf {\bibinfo {volume} {96}},\ \bibinfo
  {pages} {121203} (\bibinfo {year} {2017})}\BibitemShut {NoStop}%
\bibitem [{\citenamefont {Schryer}\ and\ \citenamefont
  {Walker}(1974)}]{SchryerJAP1974}%
  \BibitemOpen
  \bibfield  {author} {\bibinfo {author} {\bibfnamefont {N.~L.}\ \bibnamefont
  {Schryer}}\ and\ \bibinfo {author} {\bibfnamefont {L.~R.}\ \bibnamefont
  {Walker}},\ }\bibfield  {title} {\bibinfo {title} {The motion of
  180$\,^{\circ}$ domain walls in uniform dc magnetic fields},\ }\href
  {https://doi.org/http://dx.doi.org/10.1063/1.1663252} {\bibfield  {journal}
  {\bibinfo  {journal} {J. Appl. Phys.}\ }\textbf {\bibinfo {volume} {45}},\
  \bibinfo {pages} {5406} (\bibinfo {year} {1974})}\BibitemShut {NoStop}%
\bibitem [{\citenamefont {Gross}(1978)}]{GrossNPB1978}%
  \BibitemOpen
  \bibfield  {author} {\bibinfo {author} {\bibfnamefont {D.~J.}\ \bibnamefont
  {Gross}},\ }\bibfield  {title} {\bibinfo {title} {{Meron configurations in
  the two-dimensional O(3) $\sigma$-model}},\ }\href
  {https://doi.org/http://dx.doi.org/10.1016/0550-3213(78)90470-4} {\bibfield
  {journal} {\bibinfo  {journal} {Nucl. Phys.}\ }\textbf {\bibinfo {volume}
  {B132}},\ \bibinfo {pages} {439 } (\bibinfo {year} {1978})}\BibitemShut
  {NoStop}%
\bibitem [{\citenamefont {Ezawa}(2011)}]{EzawaPRB2011}%
  \BibitemOpen
  \bibfield  {author} {\bibinfo {author} {\bibfnamefont {M.}~\bibnamefont
  {Ezawa}},\ }\bibfield  {title} {\bibinfo {title} {Compact merons and
  skyrmions in thin chiral magnetic films},\ }\href
  {https://doi.org/10.1103/PhysRevB.83.100408} {\bibfield  {journal} {\bibinfo
  {journal} {Phys. Rev. B}\ }\textbf {\bibinfo {volume} {83}},\ \bibinfo
  {pages} {100408} (\bibinfo {year} {2011})}\BibitemShut {NoStop}%
\bibitem [{\citenamefont {Lin}\ \emph {et~al.}(2015)\citenamefont {Lin},
  \citenamefont {Saxena},\ and\ \citenamefont {Batista}}]{LinPRB2015}%
  \BibitemOpen
  \bibfield  {author} {\bibinfo {author} {\bibfnamefont {S.-Z.}\ \bibnamefont
  {Lin}}, \bibinfo {author} {\bibfnamefont {A.}~\bibnamefont {Saxena}},\ and\
  \bibinfo {author} {\bibfnamefont {C.~D.}\ \bibnamefont {Batista}},\
  }\bibfield  {title} {\bibinfo {title} {Skyrmion fractionalization and merons
  in chiral magnets with easy-plane anisotropy},\ }\href
  {https://doi.org/10.1103/PhysRevB.91.224407} {\bibfield  {journal} {\bibinfo
  {journal} {Phys. Rev. B}\ }\textbf {\bibinfo {volume} {91}},\ \bibinfo
  {pages} {224407} (\bibinfo {year} {2015})}\BibitemShut {NoStop}%
\bibitem [{\citenamefont {Yu}\ \emph {et~al.}(2018)\citenamefont {Yu},
  \citenamefont {Koshibae}, \citenamefont {Tokunaga}, \citenamefont {Shibata},
  \citenamefont {Taguchi}, \citenamefont {Nagaosa},\ and\ \citenamefont
  {Tokura}}]{YuNature2018}%
  \BibitemOpen
  \bibfield  {author} {\bibinfo {author} {\bibfnamefont {X.~Z.}\ \bibnamefont
  {Yu}}, \bibinfo {author} {\bibfnamefont {W.}~\bibnamefont {Koshibae}},
  \bibinfo {author} {\bibfnamefont {Y.}~\bibnamefont {Tokunaga}}, \bibinfo
  {author} {\bibfnamefont {K.}~\bibnamefont {Shibata}}, \bibinfo {author}
  {\bibfnamefont {Y.}~\bibnamefont {Taguchi}}, \bibinfo {author} {\bibfnamefont
  {N.}~\bibnamefont {Nagaosa}},\ and\ \bibinfo {author} {\bibfnamefont
  {Y.}~\bibnamefont {Tokura}},\ }\bibfield  {title} {\bibinfo {title}
  {Transformation between meron and skyrmion topological spin textures in a
  chiral magnet},\ }\href {https://doi.org/10.1038/s41586-018-0745-3}
  {\bibfield  {journal} {\bibinfo  {journal} {Nature}\ }\textbf {\bibinfo
  {volume} {564}},\ \bibinfo {pages} {95} (\bibinfo {year} {2018})}\BibitemShut
  {NoStop}%
\bibitem [{\citenamefont {Lamacraft}(2008)}]{LamacraftPRA2008}%
  \BibitemOpen
  \bibfield  {author} {\bibinfo {author} {\bibfnamefont {A.}~\bibnamefont
  {Lamacraft}},\ }\bibfield  {title} {\bibinfo {title} {Long-wavelength spin
  dynamics of ferromagnetic condensates},\ }\href
  {https://doi.org/10.1103/PhysRevA.77.063622} {\bibfield  {journal} {\bibinfo
  {journal} {Phys. Rev. A}\ }\textbf {\bibinfo {volume} {77}},\ \bibinfo
  {pages} {063622} (\bibinfo {year} {2008})}\BibitemShut {NoStop}%
\bibitem [{\citenamefont {Anderson}\ and\ \citenamefont
  {Toulouse}(1977)}]{AndersonPRL1977}%
  \BibitemOpen
  \bibfield  {author} {\bibinfo {author} {\bibfnamefont {P.~W.}\ \bibnamefont
  {Anderson}}\ and\ \bibinfo {author} {\bibfnamefont {G.}~\bibnamefont
  {Toulouse}},\ }\bibfield  {title} {\bibinfo {title} {Phase slippage without
  vortex cores: Vortex textures in superfluid $^{3}\mathrm{He}$},\ }\href
  {https://doi.org/10.1103/PhysRevLett.38.508} {\bibfield  {journal} {\bibinfo
  {journal} {Phys. Rev. Lett.}\ }\textbf {\bibinfo {volume} {38}},\ \bibinfo
  {pages} {508} (\bibinfo {year} {1977})}\BibitemShut {NoStop}%
\bibitem [{\citenamefont {Thiele}(1973)}]{ThielePRL1973}%
  \BibitemOpen
  \bibfield  {author} {\bibinfo {author} {\bibfnamefont {A.~A.}\ \bibnamefont
  {Thiele}},\ }\bibfield  {title} {\bibinfo {title} {Steady-state motion of
  magnetic domains},\ }\href {https://doi.org/10.1103/PhysRevLett.30.230}
  {\bibfield  {journal} {\bibinfo  {journal} {Phys. Rev. Lett.}\ }\textbf
  {\bibinfo {volume} {30}},\ \bibinfo {pages} {230} (\bibinfo {year}
  {1973})}\BibitemShut {NoStop}%
\bibitem [{\citenamefont {Ivanov}\ and\ \citenamefont
  {Sheka}(1994)}]{IvanovPRL1994}%
  \BibitemOpen
  \bibfield  {author} {\bibinfo {author} {\bibfnamefont {B.~A.}\ \bibnamefont
  {Ivanov}}\ and\ \bibinfo {author} {\bibfnamefont {D.~D.}\ \bibnamefont
  {Sheka}},\ }\bibfield  {title} {\bibinfo {title} {Dynamics of vortices and
  their contribution to the response functions of classical
  quasi-two-dimensional easy-plane antiferromagnet},\ }\href
  {https://doi.org/10.1103/PhysRevLett.72.404} {\bibfield  {journal} {\bibinfo
  {journal} {Phys. Rev. Lett.}\ }\textbf {\bibinfo {volume} {72}},\ \bibinfo
  {pages} {404} (\bibinfo {year} {1994})}\BibitemShut {NoStop}%
\bibitem [{\citenamefont {Tretiakov}\ \emph {et~al.}(2008)\citenamefont
  {Tretiakov}, \citenamefont {Clarke}, \citenamefont {Chern}, \citenamefont
  {Bazaliy},\ and\ \citenamefont {Tchernyshyov}}]{TretiakovPRL2008}%
  \BibitemOpen
  \bibfield  {author} {\bibinfo {author} {\bibfnamefont {O.~A.}\ \bibnamefont
  {Tretiakov}}, \bibinfo {author} {\bibfnamefont {D.}~\bibnamefont {Clarke}},
  \bibinfo {author} {\bibfnamefont {G.-W.}\ \bibnamefont {Chern}}, \bibinfo
  {author} {\bibfnamefont {Y.~B.}\ \bibnamefont {Bazaliy}},\ and\ \bibinfo
  {author} {\bibfnamefont {O.}~\bibnamefont {Tchernyshyov}},\ }\bibfield
  {title} {\bibinfo {title} {Dynamics of domain walls in magnetic nanostrips},\
  }\href {https://doi.org/10.1103/PhysRevLett.100.127204} {\bibfield  {journal}
  {\bibinfo  {journal} {Phys. Rev. Lett.}\ }\textbf {\bibinfo {volume} {100}},\
  \bibinfo {pages} {127204} (\bibinfo {year} {2008})}\BibitemShut {NoStop}%
\bibitem [{\citenamefont {Everschor-Sitte}\ and\ \citenamefont
  {Sitte}(2014)}]{Everschor-SitteJAP2014}%
  \BibitemOpen
  \bibfield  {author} {\bibinfo {author} {\bibfnamefont {K.}~\bibnamefont
  {Everschor-Sitte}}\ and\ \bibinfo {author} {\bibfnamefont {M.}~\bibnamefont
  {Sitte}},\ }\bibfield  {title} {\bibinfo {title} {Real-space berry phases:
  Skyrmion soccer (invited)},\ }\href {https://doi.org/10.1063/1.4870695}
  {\bibfield  {journal} {\bibinfo  {journal} {J. Appl. Phys.}\ }\textbf
  {\bibinfo {volume} {115}},\ \bibinfo {pages} {172602} (\bibinfo {year}
  {2014})}\BibitemShut {NoStop}%
\bibitem [{\citenamefont {Litzius}\ \emph {et~al.}(2016)\citenamefont
  {Litzius}, \citenamefont {Lemesh}, \citenamefont {Kr{\"u}ger}, \citenamefont
  {Bassirian}, \citenamefont {Caretta}, \citenamefont {Richter}, \citenamefont
  {B{\"u}ttner}, \citenamefont {Sato}, \citenamefont {Tretiakov}, \citenamefont
  {F{\"o}rster}, \citenamefont {Reeve}, \citenamefont {Weigand}, \citenamefont
  {Bykova}, \citenamefont {Stoll}, \citenamefont {Sch{\"u}tz}, \citenamefont
  {Beach},\ and\ \citenamefont {Kl{\"a}ui}}]{LitzuisNP2016}%
  \BibitemOpen
  \bibfield  {author} {\bibinfo {author} {\bibfnamefont {K.}~\bibnamefont
  {Litzius}}, \bibinfo {author} {\bibfnamefont {I.}~\bibnamefont {Lemesh}},
  \bibinfo {author} {\bibfnamefont {B.}~\bibnamefont {Kr{\"u}ger}}, \bibinfo
  {author} {\bibfnamefont {P.}~\bibnamefont {Bassirian}}, \bibinfo {author}
  {\bibfnamefont {L.}~\bibnamefont {Caretta}}, \bibinfo {author} {\bibfnamefont
  {K.}~\bibnamefont {Richter}}, \bibinfo {author} {\bibfnamefont
  {F.}~\bibnamefont {B{\"u}ttner}}, \bibinfo {author} {\bibfnamefont
  {K.}~\bibnamefont {Sato}}, \bibinfo {author} {\bibfnamefont {O.~A.}\
  \bibnamefont {Tretiakov}}, \bibinfo {author} {\bibfnamefont {J.}~\bibnamefont
  {F{\"o}rster}}, \bibinfo {author} {\bibfnamefont {R.~M.}\ \bibnamefont
  {Reeve}}, \bibinfo {author} {\bibfnamefont {M.}~\bibnamefont {Weigand}},
  \bibinfo {author} {\bibfnamefont {I.}~\bibnamefont {Bykova}}, \bibinfo
  {author} {\bibfnamefont {H.}~\bibnamefont {Stoll}}, \bibinfo {author}
  {\bibfnamefont {G.}~\bibnamefont {Sch{\"u}tz}}, \bibinfo {author}
  {\bibfnamefont {G.~S.~D.}\ \bibnamefont {Beach}},\ and\ \bibinfo {author}
  {\bibfnamefont {M.}~\bibnamefont {Kl{\"a}ui}},\ }\bibfield  {title} {\bibinfo
  {title} {Skyrmion hall effect revealed by direct time-resolved x-ray
  microscopy},\ }\href {http://dx.doi.org/10.1038/nphys4000} {\bibfield
  {journal} {\bibinfo  {journal} {Nat. Phys.}\ }\textbf {\bibinfo {volume}
  {13}},\ \bibinfo {pages} {170} (\bibinfo {year} {2016})}\BibitemShut
  {NoStop}%
\bibitem [{\citenamefont {Jiang}\ \emph {et~al.}(2017)\citenamefont {Jiang},
  \citenamefont {Zhang}, \citenamefont {Yu}, \citenamefont {Zhang},
  \citenamefont {Wang}, \citenamefont {Benjamin~Jungfleisch}, \citenamefont
  {Pearson}, \citenamefont {Cheng}, \citenamefont {Heinonen}, \citenamefont
  {Wang}, \citenamefont {Zhou}, \citenamefont {Hoffmann},\ and\ \citenamefont
  {te~Velthuis}}]{JiangNP2017}%
  \BibitemOpen
  \bibfield  {author} {\bibinfo {author} {\bibfnamefont {W.}~\bibnamefont
  {Jiang}}, \bibinfo {author} {\bibfnamefont {X.}~\bibnamefont {Zhang}},
  \bibinfo {author} {\bibfnamefont {G.}~\bibnamefont {Yu}}, \bibinfo {author}
  {\bibfnamefont {W.}~\bibnamefont {Zhang}}, \bibinfo {author} {\bibfnamefont
  {X.}~\bibnamefont {Wang}}, \bibinfo {author} {\bibfnamefont {M.}~\bibnamefont
  {Benjamin~Jungfleisch}}, \bibinfo {author} {\bibfnamefont {J.~E.}\
  \bibnamefont {Pearson}}, \bibinfo {author} {\bibfnamefont {X.}~\bibnamefont
  {Cheng}}, \bibinfo {author} {\bibfnamefont {O.}~\bibnamefont {Heinonen}},
  \bibinfo {author} {\bibfnamefont {K.~L.}\ \bibnamefont {Wang}}, \bibinfo
  {author} {\bibfnamefont {Y.}~\bibnamefont {Zhou}}, \bibinfo {author}
  {\bibfnamefont {A.}~\bibnamefont {Hoffmann}},\ and\ \bibinfo {author}
  {\bibfnamefont {S.~G.~E.}\ \bibnamefont {te~Velthuis}},\ }\bibfield  {title}
  {\bibinfo {title} {Direct observation of the skyrmion hall effect},\ }\href
  {http://dx.doi.org/10.1038/nphys3883} {\bibfield  {journal} {\bibinfo
  {journal} {Nat. Phys.}\ }\textbf {\bibinfo {volume} {13}},\ \bibinfo {pages}
  {162} (\bibinfo {year} {2017})}\BibitemShut {NoStop}%
\bibitem [{\citenamefont {Zhang}\ and\ \citenamefont
  {Li}(2004)}]{ZhangPRL2004}%
  \BibitemOpen
  \bibfield  {author} {\bibinfo {author} {\bibfnamefont {S.}~\bibnamefont
  {Zhang}}\ and\ \bibinfo {author} {\bibfnamefont {Z.}~\bibnamefont {Li}},\
  }\bibfield  {title} {\bibinfo {title} {Roles of nonequilibrium conduction
  electrons on the magnetization dynamics of ferromagnets},\ }\href
  {https://doi.org/10.1103/PhysRevLett.93.127204} {\bibfield  {journal}
  {\bibinfo  {journal} {Phys. Rev. Lett.}\ }\textbf {\bibinfo {volume} {93}},\
  \bibinfo {pages} {127204} (\bibinfo {year} {2004})}\BibitemShut {NoStop}%
\bibitem [{\citenamefont {Tatara}\ \emph {et~al.}(2008)\citenamefont {Tatara},
  \citenamefont {Kohno},\ and\ \citenamefont {Shibata}}]{TataraPR2008}%
  \BibitemOpen
  \bibfield  {author} {\bibinfo {author} {\bibfnamefont {G.}~\bibnamefont
  {Tatara}}, \bibinfo {author} {\bibfnamefont {H.}~\bibnamefont {Kohno}},\ and\
  \bibinfo {author} {\bibfnamefont {J.}~\bibnamefont {Shibata}},\ }\bibfield
  {title} {\bibinfo {title} {Microscopic approach to current-driven domain wall
  dynamics},\ }\href
  {https://doi.org/http://dx.doi.org/10.1016/j.physrep.2008.07.003} {\bibfield
  {journal} {\bibinfo  {journal} {Phys. Rep.}\ }\textbf {\bibinfo {volume}
  {468}},\ \bibinfo {pages} {213 } (\bibinfo {year} {2008})}\BibitemShut
  {NoStop}%
\end{thebibliography}%

\end{document}